\documentclass[11pt,a4paper]{article}
\usepackage{jheppub}
\usepackage{amsmath,amstext,amssymb}
\usepackage{feynmp}
\usepackage{tikz}
\usetikzlibrary{matrix,shapes,arrows,calc}
\usepackage{caption}
\usepackage{subcaption}
\usepackage{float}
\usepackage{verbatim}

\usepackage{graphics}
\usepackage{graphicx}% Include figure files
\usepackage{dcolumn}% Align table columns on decimal point
\usepackage{bm}% bold math
\usepackage{epsfig}
\usepackage{multirow}
\usepackage{dcolumn}% Align table columns on decimal point

\def\be{\begin{equation}}
\def\ee{\end{equation}}
\def\ba{\begin{eqnarray}}
\def\ea{\end{eqnarray}}

\begin{document}

\title{Analytic structure of the $n=7$ scattering amplitude in $\mathcal{N}=4$ SYM theory in multi-Regge kinematics: Conformal Regge pole contribution.}

\preprint{DESY 13-209}

\author[a]{Jochen Bartels,}%,\note{Corresponding author.}}
\author[a]{Andrey Kormilitzin,}
\author[a,b]{Lev N. Lipatov}%\note{Also at Some University.}}

\affiliation[a]{II. Institut f\"{u}r Theoretische Physik, Universit\"{a}t Hamburg, Luruper Chaussee 149,\\
D-22761 Hamburg, Germany}
\affiliation[b]{Petersburg Nuclear Physics Institute, Gatchina 188300, St.Peterburg, Russia}

% e-mail addresses: one for each author, in the same order as the authors
\emailAdd{jochen.bartels@desy.de}
\emailAdd{andrey.kormilitzin@desy.de}
\emailAdd{lipatov@mail.desy.de}

\date{\today}

%\pacs{12.38.Cy, 12.38.Aw,12.40.Vv}
\abstract{
We investigate the analytic structure of the $2\to5$ production amplitude in the planar limit of $\mathcal{N}=4$ SYM in the multi-Regge kinematics in all physical regions. We demonstrate  the close connection between Regge pole and Regge cut contributions: in a selected class of kinematic regions (Mandelstam regions) the usual factorizing Regge pole formula develops unphysical singularities that have to be absorbed and compensated by Regge cut contributions. This leads, in the corrections to the Bern-Dixon-Smirnov formula, to conformal invariant "renormalized" Regge pole expressions in the remainder function. We compute these renormalized Regge poles for the $2\to5$ production amplitude.     
}

\maketitle

\section{Introduction}

It is now well established that the Bern-Dixon-Smirnov (BDS) conjecture \cite{Bern:2005iz} for the MHV $n$-point scattering amplitude in the planar limit of the $\mathcal{N}=4$ SYM theory is incomplete for $n \ge 6$. In \cite{Alday:2007he} it has been shown that this conjecture is not correct at strong coupling and for a large number of gluons. The authors of \cite{Bartels:2008ce,Bartels:2008sc} showed that also at weak coupling this conjecture does not reproduce the correct result in different kinematic regions. Corrections to the BDS formula have been named "remainder functions", $R_{n}$, and in recent years major efforts have been made \cite{Lipatov:2009nt, Lipatov:2010qg, Lipatov:2010ad, Bartels:2010tx, Bartels:2011nz, Bartels:2011xy, Prygarin:2011gd, Bartels:2011ge, Bartels:2014ppa,Hatsuda:2014oza,Basso:2014koa, Golden:2014xqa} to determine these remainder functions. For $n=6$, the remainder function $R_{6}$ has been calculated for two, and three loops  \cite{Goncharov:2010jf,DelDuca:2009au,DelDuca:2010zg, Dixon:2011pw, Dixon:2011nj, Dixon:2012yy,Pennington:2012zj,Dixon:2013eka,Lipstein:2013xra,Golden:2013xva,DelDuca:2013lma,Caron-Huot:2013fea}. Beyond this loop expansion, it has turned out to be useful to consider a special kinematic limit, the multi-Regge limit. For the  $n=6$ point amplitude, the comparison of the BDS conjecture with the leading logarithmic approximation that extends over all orders of the coupling constant has shown that the remainder function consists of a Regge cut contribution that vanishes in the Euclidean region and in the physical region where all energies are positive. It is nonzero only in special kinematic regions, called "Mandelstam regions", which are physical regions where some of the energy variables are positive,and others are negative (the precise definition of these "mixed regions" will be given later on). These results have been generalized also beyond the leading logarithmic approximation, and there is no doubt that the multi-Regge limit plays a key role in the determination of the remainder functions.

In the comparison of the multi-Regge formula with the BDS conjecture in  \cite{Bartels:2008ce,Bartels:2008sc,Lipatov:2010qf} it was crucial to make use of the analytic structure of the $2 \to 4$ scattering amplitude in the multi-Regge limit. It is well known that in non-Abelian gauge theories the gauge bosons Reggeize, and in the leading approximation the $2 \to n+1$ production amplitudes can be written in a simple factorizing form with the exchange of Reggeized gluons in all $t$ channels. Beyond the leading approximation this factorizing form of the Regge-pole contribution remains valid in the region of all energies being positive, but the production vertices become complex-valued functions, in agreement with the results of Regge theory derived from dual models  \cite{Weis:1972ir,Weis:1972tn,Brower:1974yv} or scalar theories \cite{Drummond:1969ft}.  In \cite{Lipatov:2010qf,Bartels:2008ce} in was also shown that the simple factorized form of the Regge pole contributions is valid only in the physical region with all energy variables being positive (and also in the Euclidean region), but it takes a quite different form in all other regions, in particular in the Mandelstam regions mentioned before: in the expression for the Regge pole contribution a new term appears which contains an unphysical singularity and should be cancelled by other terms.

This representation of the Regge poles is equivalent to another representation, in which the scattering amplitude is written as a sum of $k_n$ different terms, each of them belonging to a distinct set of nonvanishing simultaneous energy discontinuities: in this representation the agreement with the Steinmann relation is explicit. For the case of $n=6$, there are five terms, i.e.  $k_6=5$; for $n>6$ the number increases rapidly: $k_7=14$, $k_8=42$ etc. As discussed in \cite{Bartels:2008ce,Bartels:2008sc}, the perturbative analysis of Yang-Mills theories shows that some of these terms contain, in addition to the Regge poles, also Regge cut singularities.
For the $2\to4$ case, this applies to two terms: in the notation of \cite{Bartels:2008ce,Bartels:2008sc}, to $W_3$ and $W_4$. In the physical region where all energies are positive, the phase factors in front $W_3$ and $W_4$ are such that the Regge cut contributions in $W_3$ and $W_4$  cancel, whereas in the Mandelstam region they add up to a nonzero result.  Both the discussions  of the Regge cut contributions and of Regge poles have made it clear that a complete analysis of the analytic structure of scattering amplitudes must include the investigation of all physical regions.  

The analysis of \cite{Lipatov:2010qf} for the $2\to 4$ amplitude has shown that there is an important connection between the Regge poles and Regge cuts which has not been seen in earlier analysis of Regge pole models \cite{Brower:1974yv}. First, it was observed that the Regge cut appears in exactly the same kinematic regions in which the Regge pole expression contains the terms with the unphysical singularities. Furthermore, both this singular Regge pole piece and the Regge cut term have the same complex phase structure: this allows us to absorb the singular Regge pole piece into the Regge cut contribution, leading to a "renormalized"  Regge pole which is free from unphysical divergences, and to a modified Regge cut definition. The existence of Regge cuts therefore resolves the problem connected with appearance of the singular pieces of the Regge poles. Conversely, without Regge cuts the standard factorizing Regge pole expression appears to be problematic.   
  
For the determination of the conformal invariant remainder function in $\mathcal{N}=4$ SYM  it is necessary to perform a careful analysis of the content of the BDS formula. In  \cite{Bartels:2008ce,Bartels:2008sc,Lipatov:2010qf} it was shown that, in multi-Regge kinematics, the BDS formula does not agree with the analytic structure outlined above in two respects: (i) the Regge pole contribution is correctly described in the region of positive energies and in the Euclidean region, but not in the Mandelstam region and (ii) in these Mandelstam regions the Regge cut contributions are contained only in the one-loop approximation, but not to all orders. This implies that the conformal invariant remainder function must (i) correct the Regge pole contribution in all kinematic regions and (ii) provide the all-loop Regge cut contribution. In view of the described interdependence between Regge pole and Regge cut  contributions, there must be a close connection between the solutions to both problems. It looks reasonable to start with the Regge pole part: here the main task is the subtraction of the singular pieces by Regge cut contributions. To be more concrete, one can attempt to use the known phase structure of the Regge pole terms in all kinematic regions to constrain the phases of the Regge cuts in such a way that they can absorb all singular terms of the Regge poles. In this subtraction,  most powerful constraints follow from the conformal invariance of the remainder function: after absorbing the singular Regge pole pieces (which by themselves are not conformal invariant) into the Regge cut contributions, the remaining "renormalized" Regge poles and the modified Regge cut terms must be conformal invariant.          

In this paper we describe this subtraction procedure for the  $2\to4$ and for $2\to 5$ cases. For the former case, most the work has been done already in earlier publications: so we only briefly review and complete our previous studies and then generalize to the $2\to 5$ case. 
In the first part (Sec. II) we analyze the general factorization formula of Regge pole contributions in all physical regions. Starting from the region of positive energies where factorization holds, we continue to other regions and derive the existence of terms with unphysical pole singularities which have to be compensated by Regge cut contributions. Particular attention will be given to the phase structure which is important in determining the phase structure of Regge cut contributions in $\mathcal{N}=4$ SYM.  We present explicit results for $2\to4$ and $2\to 5$, 
but our analysis can also be generalized to the general case $2\to n+1$. In the second part (Sec. III) we present an analysis of the BDS formula in multi-Regge kinematics in all physical regions. This analysis is general and applies 
to the case $2\to n+1$. 
 In the third part (Sec. V) we carry out the program described at the end of the previous paragraph. 
We first compute, for the case $2 \to 5$, phases of Regge cut contributions which allow to absorb the unphysical terms of the Regge poles calculated in the first part. We then define subtraction schemes for absorbing these pieces into the Regge cuts, leaving conformal invariant expressions for the Regge poles. In the final part of this section we combine these results with our findings of the BDS amplitude obtained in Sec. III, and we present predictions for the remainder function. It should be emphasized that, in this paper, we do not yet address the second part of the program, the construction of the conformal invariant Regge cut contributions. This will be left for a separate paper.    
   
\section{The Regge Pole framework}
\subsection{Factorizing Regge poles}
\begin{figure}[H]
\centering
\epsfig{file=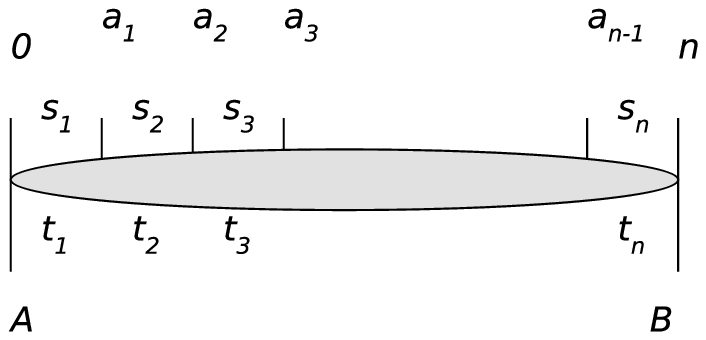}
\caption{Notations for the $2 \to n-2$ amplitude}
\label{2to_np1_lab}
\end{figure}

\begin{figure}[H]
\centering
\epsfig{file=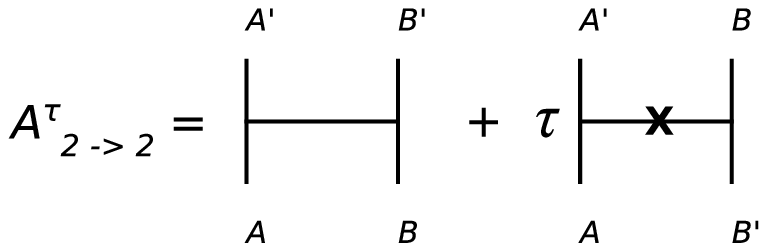,scale=0.9}
\caption{The signatured $2 \to 2$ amplitude}
\label{2to2lab}
\end{figure}

We begin with the factorized form of the fully signatured  $2 \to n+1$ production amplitude (Fig.\ref{2to_np1_lab}). The produced particles will be labeled by $a_1,...,a_{n-1}$, and they can have positive or negative energies.  
\noindent
We want to describe all physical channels of these amplitudes in the multi-Regge kinematics $s\gg |s_1|,...,|s_{n}|
\gg -t_1,...,-t_n$.
We introduce, for each $t$ channel $t_i$, the signature label $\tau_i$ which takes the values $\tau_i=+1$ or $\tau_i=-1$. For  $\tau_i=+1 (-1)$ the scattering amplitude is even (odd) under twisting the $t_i$ channel, i.e. under the crossing of the corresponding energy variables (for the simplest case, the $2 \to 2$ scattering "twisting the $t$ channel" is the same as $s \leftrightarrow u$ crossing). For our present discussion it is sufficient to consider signatured amplitudes as sums and differences of planar untwisted and twisted amplitudes. Denoting a twist by a simple cross, a signatured $2 \to 2$ scattering amplitude has the form (Fig.\ref{2to2lab})
\noindent
where the cross indicates the change of sign of the energies of the particles $B$ and $B'$. 

Generalizing this to arbitrary $n$, we write down the amplitude for the $2\to n+1$ production amplitude in the 
following form
\begin{eqnarray}
\label{generalAmp}
\frac{A^{\tau_i\tau_j...\tau_n}_{2\rightarrow n+1}}{\Gamma(t_1)\Gamma(t_n)}\;=\;|s_1|^{\omega_1}\xi_{1}V^{\tau_1\tau_2;a_1}|s_2|^{\omega_2}\xi_2 V^{\tau_2\tau_3;a_2}|s_3|^{\omega_3}\xi_3\times...\nonumber\\ \times |s_{n-1}|^{\omega_{n-1}}\xi_{n-1}V^{\tau_{n-1}\tau_{n};a_{n-1}}|s_{n}|^{\omega_{n}}\xi_{n},
\end{eqnarray}
where 
\begin{eqnarray}\label{xis}
\xi_{i}\;=\;e^{-i\pi \omega_i} - \tau_i\;\;;\;\xi_{ij}\;=\;e^{-i\pi \omega_{ij}} + \tau_i\tau_j\;\;;\;\xi_{ji}\;=\;e^{-i\pi \omega_{ji}} + \tau_i\tau_j
\end{eqnarray}
with
\be
\omega_{ij}=\omega_i-\omega_j
\ee
denote the signature factors, and 
\begin{eqnarray}
\label{prodvrx}
V^{\tau_i \tau_j;a_j}\;=\;\frac{\xi_{ij}}{\xi_i}c^{ij;a_i}_R + \frac{\xi_{ji}}{\xi_j}c^{ij;a_i}_L
\end{eqnarray}
stands for the complex-valued production vertex. 
 
As an example, for the case $2\rightarrow3$, the one particle production amplitude has a simple structure \cite{Lipatov:2010qf,Weis:1972tn}:
\begin{eqnarray}
\frac{A^{\tau_1\tau_2}_{2\rightarrow 3}}{\Gamma(t_1)|s_1|^{\omega_1}|s_2|^{\omega_2}\Gamma(t_2)}\;=\;\xi_{1}V^{\tau_1\tau_2;a_1}\xi_2\;=\;\xi_{12}\xi_2 c^{12;a_1}_R + \xi_{21}\xi_1 c^{12;a_1}_L\;\equiv\;\tilde{V}^{\tau_1\tau_2;a_1},
\label{prodvertex1}
\end{eqnarray}
where $\Gamma(t)$ is the Regge pole residue and $c_R$ and $c_L$ the Reggeon-Reggeon-particle vertices. Similarly, the production of two particles has the form \cite{Lipatov:2010qf,Weis:1972tn}  
\begin{eqnarray}
\frac{A^{\tau_1\tau_2\tau_3}_{2\rightarrow 4}}{\Gamma(t_1)|s_1|^{\omega_1}|s_2|^{\omega_2}|s_3|^{\omega_3}\Gamma(t_3)}\;=\;\xi_{1}V^{\tau_1\tau_2;a_1}\xi_2 V^{\tau_2\tau_3;a_2}\xi_3.
\end{eqnarray}
In order to arrive at a symmetric factorizing expression, we insert, for the $t_2$ channel, an additional signature factor and write
\ba\label{sixpt}
\frac{A^{\tau_1\tau_2\tau_3}_{2\rightarrow 4}}{\Gamma(t_1)|s_1|^{\omega_1}|s_2|^{\omega_2}|s_3|^{\omega_3}\Gamma(t_3)}\;= \tilde{V}^{\tau_1\tau_2;a_1}\frac{1}{\xi_2}\tilde{V}^{\tau_2\tau_3;a_2}, 
\ea
where
\be
\tilde{V}^{\tau_1\tau_2;a_1}= \xi_1 V^{\tau_1\tau_2;a_1}\xi_2.
\ee
Generalizing to the case $2 \to n+1$, we see that for each "inner" $t_i$ channel, $t_2,...,t_{n-1}$, we need an extra "propagator" $1/\xi_i$. With this rule Eq.\eqref{generalAmp} can be written in the convenient form
\ba
\label{generalAmpfact}
\frac{A^{\tau_i\tau_j...\tau_n}_{2\rightarrow n+1}}{\Gamma(t_1) |s_1|^{\omega_1} |s_2|^{\omega_2}...
|s_{n}|^{\omega_{n}}\Gamma(t_n)} 
\;=\; \tilde{V}^{\tau_1\tau_2;a_1} \frac{1}{\xi_2} \tilde{V}^{\tau_2\tau_3;a_2} \frac{1}{\xi_3} ...
\frac{1}{\xi_{n-1}} \tilde{V}^{\tau_{n-1}\tau_{n};a_{n-1}}.
\ea
It will be useful to write this formula as an expansion in monomials of signatures $\tau_i$. In such an expansion, terms without any $\tau_i$ can be identified as the planar approximation in the kinematic region where all energies are positive. For the case of $n=6$, terms proportional to $\tau_1 \tau_3$ correspond to the planar amplitude where the particles $a_1$ and $a_2$ have become incoming: this is one of the Mandelstam regions where, according to  the analysis in \cite{Bartels:2008ce,Bartels:2008sc}, the Regge cut contribution will appear. 
 
In order to obtain this representation we observe that the production vertex, Eq.\eqref{prodvertex1}, can be expanded as:
\ba
\tilde{V}^{\tau_1\tau_2;a}= e^{-i\pi \omega_1} c^{12;a_1}_R + e^{-i\pi \omega_2} c^{12;a_1}_L  
-\tau_1 e^{-i\pi \omega_1} \left( e^{-i\pi \omega_1} c^{12;a_1}_R + e^{-i\pi \omega_2} c^{12;a_1}_L  \right)\nonumber\\
-\tau_2 e^{-i\pi \omega_2} \left( e^{-i\pi \omega_1} c^{12;a_1}_R + e^{-i\pi \omega_2} c^{12;a_1}_L  \right)
+ \tau_1\tau_2 \left( e^{-i\pi \omega_2} c^{12;a_1}_R + e^{-i\pi \omega_1} c^{12;a_1}_L \right),
\label{vertextauexp}
\ea
and the propagator can be written in the form
\begin{eqnarray}
\label{pseudopropagator}
\frac{1}{\xi_2}\;=\;\frac{1}{e^{-i\pi\omega_2}-\tau_2}\;=\;\frac{e^{-i\pi\omega_2}+\tau_2}{-2i\sin(\pi\omega_2)e^{-i\pi\omega_2}}.
\end{eqnarray}
Note the appearance of the nonphysical poles $\sim 1/\sin(\pi \omega_2)$ which should be cancelled by the Regge cut contributions.
     
With these ingredients it is straightforward to find the expansion in monomials for the general $2\to n+1$ amplitude.

%%%%%%%%%%%%%%%%%%%%%%%%%%%%%%%%%%%%%%%%%%%%%%%%%%%%%%%%%%%%%%%%%%%%%%%%%%%%%%%%%%%%%%%%%%%%%%%%%%%
%%%%%%%%%%%%%%%%%%%%%%%%%%%%%%%%%%%%%%%%%%%%%%%%%%%%%%%%%%%%%%%%%%%%%%%%%%%%%%%%%%%%%%%%%%%%%%%%%%%

\subsection{Generating Function approach for the Regge Pole formula}

To be definite, let us from now on concentrate on planar $\mathcal{N}=4$ SYM. It will be convenient to define a generating function for the pole-term coefficients. Let us briefly introduce the idea behind it. We are interested in the analytical continuation of the planar  scattering amplitude to arbitrary kinematic regions in multi-Regge kinematics. During such a continuation, various factors and phases may appear. As explained above, each particular kinematic region can be reached by a sequence of twists (crosses) of $t$ channels, and each such twist is denoted by a corresponding factor $\tau$\footnote{It should be clear that, from now on, $\tau$ is no longer related to signature but simply denotes kinematic regions}. Thus, it is instructive to have a list of all possible phases and factors that appear due to continuation for each appropriate kinematic configuration. One may also think of a different point of view on the scattering amplitude. Instead of having one analytical function of kinematic invariants and then continuing to arbitrary physical and nonphysical kinematic regions, one can introduce  a generating function, $P_{2 \to n}$, which is given as a sum of amplitudes in all physical regions.  As a simple example, consider such a  generating function of the $2 \rightarrow 3$ scattering process Fig.\ref{2to3lab}:
\begin{figure}[H]
\centering
\epsfig{file=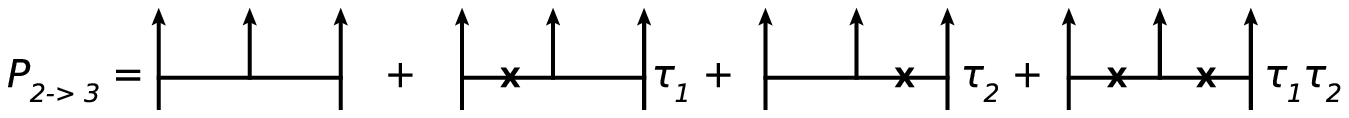}
\caption{The generating function for the $2\rightarrow 3$ production process written in terms of monomials of $\tau_1$, $\tau_2$.}
\label{2to3lab}
\end{figure}
\noindent
Turning now to the BDS formula, applied to the $2\to3$ amplitude \cite{Bartels:2008ce}, we have  for the Reggeon vertices in (\ref{prodvrx}): 
\begin{eqnarray}\label{vertexRL}
c^{ii+1;a}_R\;=\;|\Gamma_{i,i+1}|\frac{\sin(\pi\omega_i-\pi\omega_a)}{\sin(\pi\omega_i-\pi\omega_{i+1})}\;\;\;\;;\;\;\;c^{ii+1;a}_L\;=\;|\Gamma_{i,i+1;a}|\frac{\sin(\pi\omega_{i+1}-\pi\omega_a)}{\sin(\pi\omega_{i+1}-\pi\omega_i)}.
\end{eqnarray}
Here $i$ labels the $t$ channel (for the $2\to3$ case we have $i=1$ only), $a$ denotes the produced particle. Going to the physical region where all energies are positive, this allows to write the Reggeon-Reggeon-Gluon vertex 
$\Gamma_{i,j;a}$ (see Eqs.(19)-(22) \cite{Lipatov:2010qf}) in the form:
\begin{eqnarray}
\label{omegas}
\Gamma_{i,i+1;a}(\ln(\kappa_{a}-i\pi))\;=\;|\Gamma_{i,i+1;a}|e^{i\pi\omega_{a}}.
\end{eqnarray}
Here the expansions in powers of $a=\frac{g^2N_c}{8\pi^2}$ are given by:
\begin{eqnarray}
\omega_i\;=\;-\frac{\gamma_K}{4}\ln\frac{|q_i|^2}{\lambda^2}\;,\;\;\gamma_K=4a + {\it{O}}(a^2),
\end{eqnarray}
where $\gamma_K$ is the cusp anomalous dimension and $\lambda^2\,\equiv\,\mu^2e^{1/\epsilon}$ for $D=4-2\epsilon$ with $\epsilon\rightarrow 0^{-}$,
\begin{eqnarray}
\label{omega-a}
\omega_a\;=\;-\frac{\gamma_K}{8}\ln\frac{|q_i|^2|q_{i+1}|^2}{|k_{a_{i+1}}|^2\lambda^2}
\end{eqnarray}
with $k_{a_{i+1}} = q_i-q_{i+1}$, 
and 
\begin{eqnarray}
\ln|\Gamma_{i,i+1}|\;=\;\frac{\gamma_K}{4}\left(-\frac{1}{4}\ln^2\frac{|q_i-q_{i+1}|^2}{\lambda^2}-\frac{1}{4}\ln^2\frac{|q_i|^2}{|q_{i+1}|^2}+\frac{1}{2}\ln\frac{|q_i|^2|q_{i+1}|^2}{\lambda^4}\ln\frac{|q_i-q_{i+1}|^2}{\mu^2}+\frac{5}{4}\zeta(2)\right).\nonumber\\
\end{eqnarray}
Let us now return to the generating functions $P_{2 \to n}$, to the sum of amplitudes in all kinematic regions. It is convenient to divide by factors which are common to all kinematic regions. Beginning with the case $2\to3$, Namely, using the explicit form Eq.\eqref{prodvertex1} with Eq.\eqref{vertexRL} one arrives at
\begin{eqnarray}
\label{fivepointpartitionfunction}
P_{2\to3}&=&
\frac{A_{2\to3}}{\Gamma(t_1)|s_1|^{\omega_1}||\Gamma_{1,2}| |s_2|^{\omega_2}\Gamma(t_2)}\;\nonumber\\
&=&\tilde{V}_{red}^{\tau_1\tau_2;a}\\ &=&\;\;e^{-i\pi\left(\omega_1 + \omega_2 - \omega_a\right)} - e^{-i\pi\left(\omega_2 - \omega_a\right)}\tau_1 - e^{-i\pi\left(\omega_1 - \omega_a\right)}\tau_2  + e^{- i\pi\omega_a}\tau_1\tau_2.\nonumber
\end{eqnarray}
Here we have defined a reduced vertex by
\be
\tilde{V}_{red}^{\tau_1\tau_2;a}  =\frac{\tilde{V}^{\tau_1\tau_2;a}}{|\Gamma_{1,2}|}\;=\;e^{-i\pi\left(\omega_1 + \omega_2 - \omega_a\right)} - e^{-i\pi\left(\omega_2 - \omega_a\right)}\tau_1 - e^{-i\pi\left(\omega_1 - \omega_a\right)}\tau_2  + e^{- i\pi\omega_a}\tau_1\tau_2,
\ee
which consists of phases only. 

As the next example we calculate, from  Eq.\eqref{sixpt}, the six-point generating function   
(cf.(\cite{Lipatov:2010qf}):
\begin{eqnarray}
\label{sixpointpartitionfunction}
P_{2\rightarrow 4}&=&
\frac{A_{2\rightarrow 4}}{\Gamma(t_1)|s_1|^{\omega_1}|\Gamma_{1,2}| |s_2|^{\omega_2}
|\Gamma_{2,3}|
|s_3|^{\omega_3}\Gamma(t_3)}\nonumber \\
&=&\;\tilde{V}_{red}^{\tau_1\tau_2;a}\frac{1}{\xi_2}\tilde{V}_{red}^{\tau_2\tau_3;b}\nonumber\\ \;&=&e^{-i\pi(\omega_1+\omega_2+\omega_3-\omega_a-\omega_b)}\nonumber\\
&-&e^{-i\pi(\omega_2+\omega_3-\omega_a-\omega_b)}\tau_1 -  e^{-i\pi(\omega_1+\omega_3-\omega_a-\omega_b)}\tau_2 - e^{-i\pi(\omega_1+\omega_2-\omega_a-\omega_b)}\tau_3 \nonumber\\&+& e^{-i\pi(\omega_3 + \omega_a-\omega_b)}\tau_1\tau_2 + e^{-i\pi(\omega_1 - \omega_a + \omega_b)}\tau_2\tau_3  \nonumber \\ &+& e^{-i\pi\omega_2}\left\{\cos(\pi\omega_{ab}) + i\left(\sin(\pi\omega_a+\pi\omega_b) - 2e^{i\pi\omega_2}\frac{ \sin (\pi\omega_a) \sin (\pi\omega_b)}{\sin(\pi\omega_2)} \right)\right\}\tau_1\tau_3 + \nonumber \\ &-&  \left\{\cos(\pi\omega_{ab}) - i\left(\sin(\pi\omega_a+\pi\omega_b)  -2e^{-i\pi\omega_2}\frac{\sin(\pi\omega_a) \sin(\pi\omega_b)}{\sin(\pi\omega_2)}\right)\right\} \tau_1\tau_2\tau_3,\nonumber\\
\end{eqnarray}
where $\omega_{ab}\;=\;\omega_a - \omega_b$. The careful reader may notice that this expression has a mirror symmetry with respect to right and left ($a \leftrightarrow b$) exchange. This fact will be important in the future. 

Concluding this part, on can write a general expression for the generating function for an arbitrary number of produced particles $2\rightarrow n+1$:
\ba
\label{generalpartitionfunction}
P_{2\rightarrow n+1}&=&\frac{A_{2\rightarrow n+1}}{\Gamma(t_1) |s_1|^{\omega_1} |\Gamma_{1,2}| |s_2|^{\omega_2}...
|\Gamma_{n-1,n}| |s_{n}|^{\omega_{n}}\Gamma(t_n)} \nonumber \\
&=&\; \tilde{V}_{red}^{\tau_1\tau_2;a_1} \frac{1}{\xi_2} \tilde{V}_{red}^{\tau_2\tau_3;a_2} \frac{1}{\xi_3} ...
\frac{1}{\xi_{n-1}} \tilde{V}_{red}^{\tau_{n-1}\tau_{n};a_{n-1}}\nonumber\\
&=&\;a_0 + a_1\tau_1 + a_2\tau_2 + a_{12}\tau_1\tau_2 + ... + a_{1..n}\tau_{1}...\tau_n.
\ea
The r.h.s. can be written as a polynomial in the $\tau_i$, and the coefficients consist of phases and trigonometric functions. In the Appendix A we list, for the cases $2\to3$, $2\to4$, and $2\to5$,  all coefficients of the generating function.

%%%%%%%%%%%%%%%%%%%%%%%%%%%%%%%%%%%%%%%%%%%%%%%%%%%%%%%%%%%%%%%%%%%%%%%%%%%%%%%%%%%%%%%%%%%%%%%%%%%%%%
%%%%%%%%%%%%%%%%%%%%%%%%%%%%%%%%%%%%%%%%%%%%%%%%%%%%%%%%%%%%%%%%%%%%%%%%%%%%%%%%%%%%%%%%%%%%%%%%%%%%%%

\subsection{Rules: a few particular cases}

\begin{figure}[H]
\centering
\epsfig{file=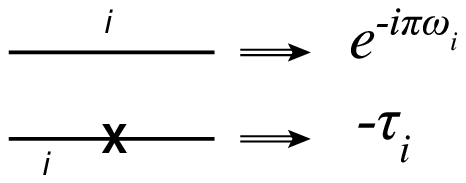}
\caption{Two types of propagators in channel $i$.}
\label{propagators}
\end{figure}

It will be useful to extract, from the particular cases given above, a few general rules. Let us begin with the case $n=5$. As we have said before, the term without any $\tau$  belongs to the planar amplitude in the physical region with all positive energies. On the rhs of Eq.\eqref{fivepointpartitionfunction} we have:
\be
 e^{i\pi\omega_a}e^{-i\pi\left(\omega_1+\omega_2\right)}.
 \ee
As expected, the amplitude has the simple factorized form, with phase factors for the produced particle, $e^{i\pi \omega_a}$, and for the exchange channels, $e^{-i\pi \omega_1}$ and $e^{-i\pi \omega_2}$. 
As to the remaining three terms for $n=5$ we observe the following pattern: each $t$ channel without a twist comes with a phase factor  
$e^{-i\pi \omega_i}$, each $t$ channel with a twist carries the factor $-1$: 
\begin{itemize}
\item twisted propagator: $\rightarrow$ $-1$
\item untwisted propagator in channel $t_i$: $\rightarrow$ $e^{-i\pi\omega_i}$
\end{itemize}
An illustration is given in Fig.\ref{propagators}
\noindent
Turning to $n=6$, all but two terms are of the form that we have just described:
phase factors for the propagators and for the production vertices. It is important to note that in all these terms the pole $\sim 1/\sin (\pi \omega_2)$ from the propagator of the $t_2$ channel cancels. New features appear for $\tau_1 \tau_3$ and $\tau_1 \tau_2 \tau_3$, namely terms where the poles $\sim 1/\sin (\pi \omega_2)$ from the propagator Eq.(\ref{pseudopropagator}) remain.  The term proportional to $\tau_1 \tau_3$  belongs to the planar amplitude continued into the physical regions where particles $a$ and $b$ are incoming Fig.\ref{Fig_2to4_twist13}: 
\begin{figure}[H]
\centering
\epsfig{file=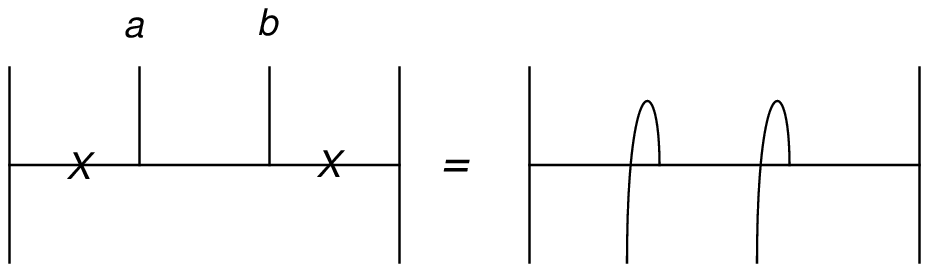}
\caption{Illustration of the term $\tau_1\tau_3$.}
\label{Fig_2to4_twist13}
\end{figure}
\noindent
This kinematic region is the one in which the Regge cut appears \cite{Bartels:2008ce,Bartels:2008sc}.
For this term we find from the rhs of Eq.\eqref{sixpointpartitionfunction}
\be
\label{2to4twist13a}
 = e^{-i\pi \omega_2} \Big[ \cos (\pi(\omega_a-\omega_b)) + i \sin ( \pi(\omega_a+\omega_b)) - 2i 
\frac{\cos (\pi \omega_2) \sin (\pi \omega_a) \sin (\pi \omega_b)}{\sin (\pi \omega_2)} \Big].
\ee
which we rewrite as
\be
\label{2to4twist13}
\text{Eq.}\eqref{2to4twist13a} =e^{-i\pi\omega_2}\left[e^{i\pi\left(\omega_a+\omega_b\right)} - 2ie^{i\pi\omega_2}\frac{\sin(\pi\omega_a)\sin(\pi\omega_b)}{\sin(\pi\omega_2)} \right] .
\ee
Here the first term is of the same form as discussed before, whereas the second term is new: it has an unphysical pole in $\sin (\pi \omega_2)$.

The important observation made in \cite{Lipatov:2010qf} is that the last two terms can be included in the Regge cut contribution, because they have the same phase structure as the Regge cut. 
This is the simplest example of the general feature that a Regge pole amplitude which, for positive energies, has the factorizing form, after analytic continuation, exhibits unphysical poles (in our case:  $\sim 1/\sin (\pi \omega_2)$). From \cite{Bartels:2008ce,Bartels:2008sc} we know that, in Yang-Mills theories, the $2\to4$ amplitude contains a Regge cut contribution with the same phase $i e^{-i\pi \omega_2}$, which can absorb the singular piece in Eq.\eqref{2to4twist13a} of the Regge pole contribution. 

An analogous discussion applies also to the term proportional to $\tau_1\tau_2\tau_3$. Note, however, that in this case the first term (see Appendix A) is of the form 
\be
-e^{-i\pi(\omega_a+\omega_b)}.
\ee  
As expected, there are no phases from  $t$ channel propagators, but for the production vertices we have 
$e^{-i\pi \omega_a}$ instead of  $e^{i\pi \omega_a}$.

Moving on to $n=7$, we again note the appearance of pole terms: the coefficient of $\tau_1\tau_3$ is illustrated in Fig.\ref{Fig_2to5_twist13}:
\begin{figure}[H]
\centering
\epsfig{file=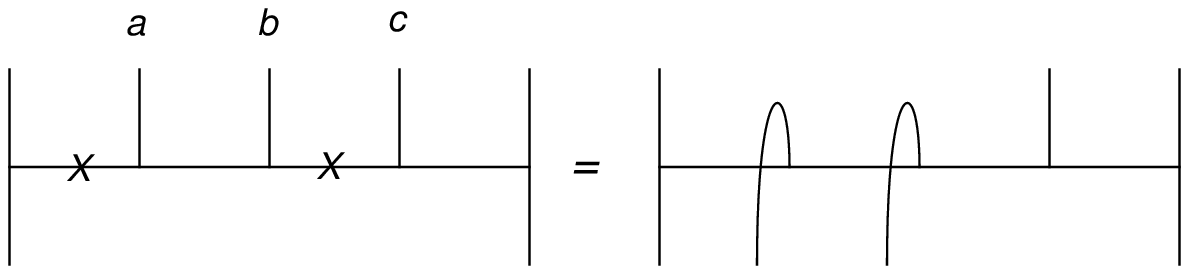}
\caption{Illustration of the term $\tau_1\tau_3$.}
\label{Fig_2to5_twist13}
\end{figure}
\noindent
It has the form (Appendix A)
\be
e^{-i\pi\left(\omega_2+\omega_4\right)}e^{i\pi\omega_c}\left[e^{i\pi\left(\omega_a+\omega_b\right)} - 2ie^{i\pi\omega_2}\frac{\sin(\pi\omega_a)\sin(\pi\omega_b)}{\sin(\pi\omega_2)} \right].
\ee
It is easily obtained from the analogous term of the $2\to4$ amplitude by multiplication with 
$e^{i \pi \omega_c}$ (for the additional vertex of particle $c$) and by $e^{-i \pi \omega_4}$ (for the untwisted 
propagator of the $t_4$ channel). The pole term $\sim 1/\sin (\pi \omega_2)$ belongs to the $t_2$ channel, and later on we will show that it can be combined with the Regge cut contribution in the same $t$ channel. An analogous discussion holds for the coefficient of $\tau_2 \tau_4$. Next let us consider the coefficient of  $\tau_1 \tau_4$ (Fig.\ref{Fig_2to5_twist14}):
\begin{figure}[H]
\centering
\epsfig{file=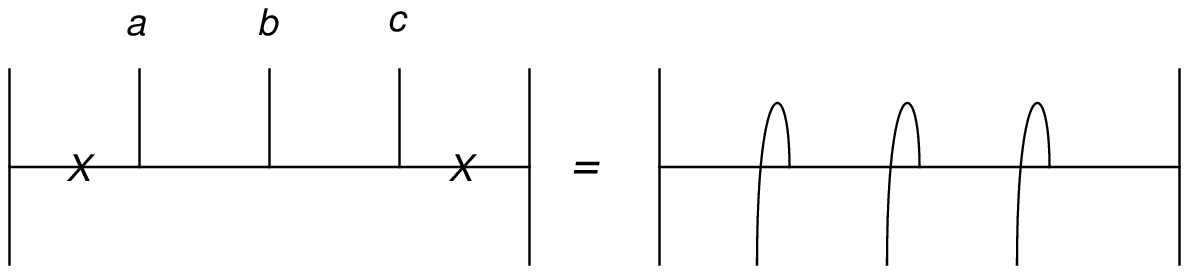}
\caption{Illustration of the term $\tau_1\tau_4$.}
\label{Fig_2to5_twist14}
\end{figure}
\noindent
The corresponding term on the rhs of Eq.\eqref{generalpartitionfunction} is (see Appendix A)
\be
\left[e^{i\pi\left(\omega_a+\omega_b+\omega_c\right)}e^{-i\pi\left(\omega_2+\omega_3\right)} - 2i\frac{\sin(\pi\omega_a)\sin(\pi\omega_b)\sin(\pi\omega_c)}{\sin(\pi\omega_2)\sin(\pi\omega_3)}\right].
\ee
Again, the first term is of the same form as the cases discussed above, whereas the double pole term belongs to the 
$t_2$ and $t_3$ channels and has to be combined with the Rege cut contribution extending over these two channels. Finally, we look at the coefficient of $\tau_1 \tau_2 \tau_3 \tau_4$ (Fig.\ref{Fig_2to5_twist1234})
\begin{figure}[H]
\centering
\epsfig{file=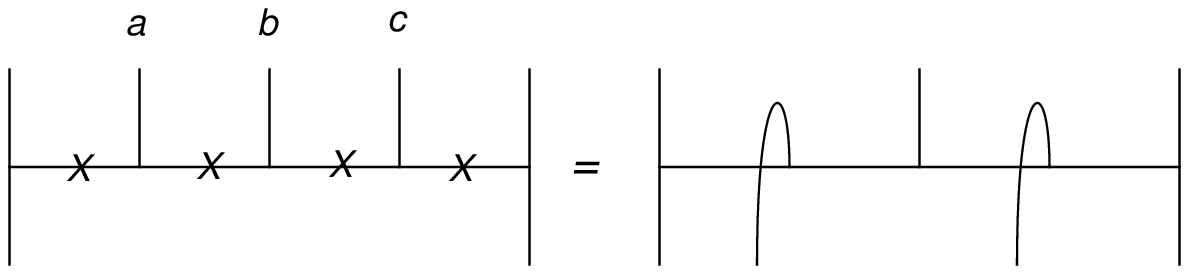}
\caption{Illustration of the term $\tau_1\tau_2 \tau_3\tau_4$.}
\label{Fig_2to5_twist1234}
\end{figure}
\noindent
It has the form
\be
\left[e^{i\pi\left(\omega_b-\omega_a-\omega_c\right)} - 2i\frac{\sin(\pi\omega_{2a})\sin(\pi\omega_b)\sin(\pi\omega_{3c})}{\sin(\pi\omega_2)\sin(\pi\omega_3)} \right],
\ee
and there is again a double pole which has to absorbed by the Rege cut contribution extending over the $t_2$ and $t_3$ channels. The first term deviates from the previous cases: for the production vertex of particle $b$ we have $e^{i\pi \omega_b}$, whereas particles $a$ and $c$ become the complex conjugate.   

In Appendix A we present, for the cases $2\to3$ (Table~\ref{table1}), $2\to4$ (Table~\ref{table2}),and $2\to5$ (Table~\ref{table3}), a complete list of all coefficients of the generating function. In all cases we first find a term with a pure phase. For  the "generalized Mandelstam regions", there are, in addition, terms with simple, double, and multipoles of the form   $\sim 1/\left( \sin (\pi \omega_i) \sin (\pi \omega_j) ...\sin(\pi \omega_k)\right) $. A closer inspection shows a one-to-one correspondence between these singular terms and Regge cut contributions: we will explicitly study the case $n=7$ and show that these Regge cut pieces can be used to absorb all singular terms. 
%%%%%%%%%%%%%%%%%%%%%%%%%%%%%%%%%%%%%%%%%%%%%%%%%%%%%%%%%%%%%%%%%%%%%%%%%%%%%%%%%%%%%%%%%%%%%%%%%%%%%%%%%%%%%%%%%%%%%%%%

\subsection{The general case: recurrence  relations}
In order to analyze the structure for the general case  it is useful to make use of recurrence relations. To begin with, consider the generating function of the five point amplitude, $P_{2\to3}^{\tau_1\tau_2}$ (Eq.\eqref{fivepointpartitionfunction}). Due to the factorization property Eq.\eqref{generalpartitionfunction}, we can obtain the six-point generating function by applying a  recurrence operator $\tilde{K}$. 
\be
P_{2\to4}  = \tilde{V}_{red}^{\tau_1,\tau_2;a}\frac{1}{\xi_{2}} \tilde{V}_{red}^{\tau_2,\tau_3;b}=
\tilde{V}_{red}^{\tau_1,\tau_2;a} \tilde{K}(\tau_2,\tau_3;b)
\ee
with
\be
\tilde{K}(\tau_2,\tau_3;b)\;=\;\frac{1}{\xi_{2}} \tilde{V}_{red}^{\tau_2,\tau_3}.
\ee
Explicitly:
\begin{eqnarray}\label{evolvop}
\tilde{K}(\tau_2,\tau_3;b)\;=\;e^{-i\pi\left(\omega_3-\omega_b\right)}-\frac{\sin(\pi\omega_{2b})}{\sin(\pi\omega_2)}\tau_3+\frac{\sin(\pi\omega_b)}{\sin(\pi\omega_2)}\tau_2\tau_3.
\end{eqnarray}
Note that $\tilde{K}$ is not symmetric with respect to the monomial representation. In particular, it does not contain a term proportional to $\tau_2$. Nevertheless, the resulting generating function, $P_{2\to n+1}$,  
\begin{eqnarray}
P_{2 \to n+1}=\tilde{V}_{red}^{\tau_1,\tau_2;a_1}\;\tilde{K}(\tau_{2},\tau_{3};a_2)...\tilde{K}(\tau_{n-1}.\tau_{n};a_{n-1})
\label{P_red-general}
\end{eqnarray}
is symmetric.

\begin{figure}[H]
\centering
\epsfig{file=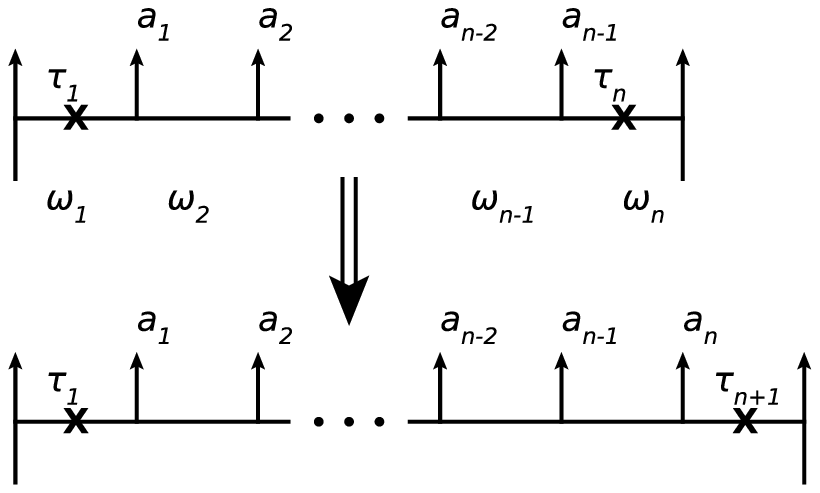}
\caption{Initial configuration $\tau_1\tau_n$.}
\label{config_1n}
\end{figure}

In Appendix B we present a more general discussion of the coefficients of different configuration of $\tau$'s. 
Here we only discuss one special case which corresponds to two crosses in the first (left) and in the last (right) channel Fig.\ref{config_1n}. As before, we consider the case $2 \to n+1$  with $n$ $t$ channels ($t_1,...,t_n$) and ($\omega_1,...\omega_n$), and $n-1$ produced particles labeled by ${a_1}, ..., {a_{n-1}}$. 
\noindent
and we want to prove, by induction, that the coefficient of $\tau_1\tau_{n}$ in $P_{2\to n+1}$ is given by
\begin{eqnarray}
\label{t1tn}
\left\{e^{-i\pi\left(\omega_2+\omega_3+...\omega_{n-1}\right)}e^{i\pi\left(\omega_{a_1}+\omega_{a_2}+...+\omega_{a_{n-1}}\right)} \right. \nonumber \\ \left.
-2i\frac{\sin(\pi\omega_{a_1})\sin(\pi\omega_{a_2})...\sin(\pi\omega_{a_{n-1}})}{\sin(\pi\omega_{2})\sin(\pi\omega_{3})...\sin(\pi\omega_{n-1})} \right\}\tau_1\tau_n.
\end{eqnarray}
For this we also need to show that the coefficient proportional to $\tau_1$ is
\begin{eqnarray}
\left\{e^{-i\pi\left(\omega_2+\omega_3+...\omega_{n}\right)}e^{i\pi\left(\omega_{a_1}+\omega_{a_2}+...+\omega_{a_{n-1}}\right)}\right\}\tau_1.
\label{tau1inP2ton}
\end{eqnarray}
To begin with the simplest case, $2 \to4$, we have for the coefficient $\tau_1\tau_3$ (Eq.\eqref{2to4twist13} or Appendix A). 
\begin{eqnarray}
\label{t1t3}
 e^{-i\pi \omega_2}  e^{i\pi (\omega_{a_1}+\omega_{a_2})} 
-2i\frac{\sin(\pi\omega_{a_1}) \sin(\pi\omega_{a_2})}{\sin(\pi\omega_{2})},
\end{eqnarray}
whereas the coefficient of $\tau_1$ is
\begin{eqnarray}
e^{-i\pi (\omega_2+\omega_3)} e^{i\pi\left(\omega_{a_1}+\omega_{a_2} \right)}.
\label{tau1inP2to4}
\end{eqnarray}
Let us now prove, by induction, our assertion. In order to go from the case $2\to n+1$ to the case $2 \to n+2$, we multiply $P_{2\to n+1}$  with the kernel $\tilde{K}(\tau_ n,\tau_{n+1};a_n)$
\begin{eqnarray}\label{operatoTNNP1}
\tilde{K}(\tau_ n,\tau_{n+1};a_n)\;=\;e^{-i\pi\omega_{n+1}}e^{i\pi\omega_{a_{n}}}-\frac{\sin(\pi\omega_n-\pi\omega_{a_{n}})}{\sin(\pi\omega_{n})}\tau_{n+1}+\frac{\sin(\pi\omega_{a_n})}{\sin(\pi\omega_n)}\tau_n\tau_{n+1}.
\end{eqnarray}
Within this product, the relevant terms  are 
\be
P_{2\to n+1} \cdot \tilde{K}(\tau_ n,\tau_{n+1};a_n)=\\
\Big[\{...\}\tau_1 + \{...\}\boldsymbol{\tau_{1}\tau_{n}}\Big]
\cdot \Big[1\{...\}+   \{...\}\boldsymbol{\tau_{n+1}}+\{...\}\tau_{1} \tau_{n+1} \Big],
\ee
where, by assumption, in the first square bracket we use Eq.\eqref{t1tn} and Eq.\eqref{tau1inP2ton}, and the second bracket is given in Eq.\eqref{operatoTNNP1}.

We immediately see that, on the rhs, the coefficient of $\tau_1$ comes from the product of the first terms 
in each  square bracket and equals
\ba
e^{-i\pi (\omega_2+...+\omega_{n+1})}e^{i\pi\left(\omega_{a_1}+...+\omega_{a_n} \right)}.
\ea
This proves the second part of our assertion. Next, in order to calculate, the contribution proportional to 
$\tau_1\tau_{n+1}$, one should take into account two terms: 
the product of the term $\sim \tau_1 \tau_n$ in the first bracket with the term $\tau_n\tau_{n+1}$ in 
the second  bracket, and the product of the term $\tau_1$  in the first 
bracket with the term $\tau_{n+1}$ in the second bracket. 
When combining these two contributions, the following identity is useful:
\begin{eqnarray}\label{identity}
\frac{\sin(\pi\omega_{a_n})}{\sin(\pi\omega_n)}+e^{-i\pi\omega_n}\frac{\sin(\pi\omega_n-\pi\omega_{a_n})}{\sin(\pi\omega_n)}\;=\;e^{-i\pi\omega_n}e^{i\pi\omega_{a_n}}.
\end{eqnarray}
One arrives at
\begin{eqnarray}\label{t1tn+1}
\left\{e^{-i\pi\left(\omega_2+\omega_3+...\omega_{n}\right)}e^{i\pi\left(\omega_{a_1}+\omega_{a_2}+...+\omega_{a_{n}}\right)} \right. \nonumber  \left.
-2i\frac{\sin(\pi\omega_{a_1})\sin(\pi\omega_{a_2})...\sin(\pi\omega_{a_{n}})}{\sin(\pi\omega_{2})\sin(\pi\omega_{3})...\sin(\pi\omega_{n})} \right\}\tau_1\tau_{n+1},
\end{eqnarray} 
which proves the first part of our assertion. 

Concluding this part, according to Eq.\eqref{P_red-general}, each coefficient of the $\tau$ expansion in Eq.\eqref{generalpartitionfunction} can be calculated recursively, by multiplying the iterative kernel Eq.\eqref{operatoTNNP1} with  the initial expression $\tilde{V}_{red}^{\tau_1,\tau_2;a_1}$.\footnote{ Although it is possible to calculate each coefficient in the expansion by using these recurrence relations, practically it is more efficient to use simple code with Wolfram Mathematica, which generates these coefficients immediately. The simplest implementation might be iterative multiplication with the kernel $\tilde{K}$.}

%application of the operator $\tilde{K}$ Eq.\eqref{evolvop} with appropriate $\tau_n, \tau_{n+1}, \omega_n$ on the initial state Eq.\eqref{inicond}.}

%%%%%%%%%%%%%%%%%%%%%%%%%%%%%%%%%%%%%%%%%%%%%%%%%%%%%%%%%%%%%%%%%%%%%%%%%%%%%%%%%%%%%%%%%%%%%%%%%%%%%%%%%%%%%%%%%%%%%%%%%%%%
%%%%%%%%%%%%%%%%%%%%%%%%%%%%%%%%%%%%%%%%%%%%%%%%%%%%%%%%%%%%%%%%%%%%%%%%%%%%%%%%%%%%%%%%%%%%%%%%%%%%%%%%%%%%%%%%%%%%%%%%%%%%

\section{Generating function for the BDS amplitudes in the multi-Regge kinematics}

\subsection{Motivation}

\begin{figure}[H]
\centering
\epsfig{file=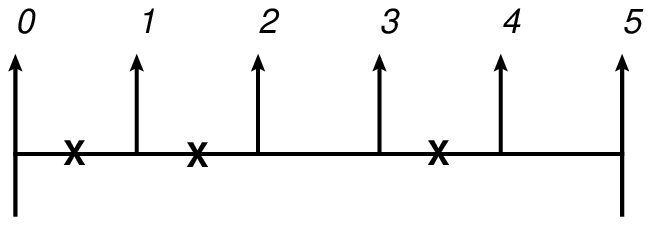}
\caption{Example of a diagram with twists in the channels 1, 2, and 4. In Eq.\eqref{generalexapnsion} it corresponds to the term $\tau_1\tau_2\tau_4$.}
\label{fig:fig10}
\end{figure}

In order to determine the remainder function in each physical region for the pole and cut combinations, let us now find the phase structure of the BDS amplitude \cite{Bern:2005iz} in the different kinematic regions. Again, we find it convenient to define a generating function:
\begin{eqnarray}\label{generalexapnsion}
A_{BDS}\;=\;a_0 + a_1\tau_1 + a_2\tau_2 +...+a_n\tau_n + a_{12}\tau_1\tau_2 + a_{13}\tau_1\tau_3+...a_{1..n}\tau_1\tau_2...\tau_n.
\end{eqnarray}
In this expansion, each monomial of the twists $\tau_i...\tau_j$  defines a kinematic region, and the coefficient
$a_{i...j}$ is the BDS prediction for this region.
As before, each term in the expansion corresponds to a diagram of the type shown in Fig. \ref{fig:fig10}.
\noindent
The following discussion of the BDS formula will be similar to the previous study of the Regge pole model, but 
the results all be quite different.

\begin{figure}[H]
\centering
\epsfig{file=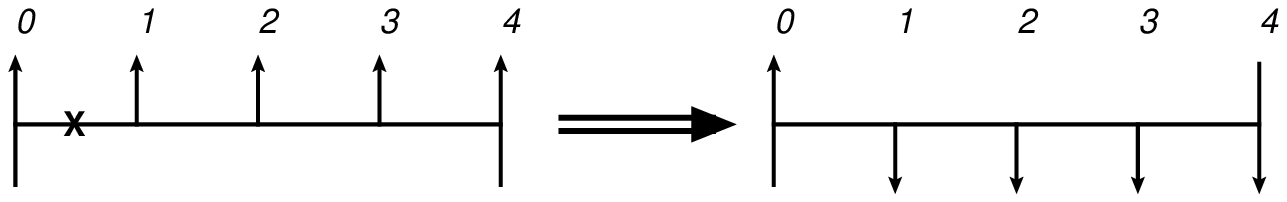}
\caption{Example of diagram with a twist in channel 1, which in the expansion Eq.\eqref{generalexapnsion} corresponds to the term $\tau_1$.}
\label{firsttwist}
\end{figure}

\begin{figure}[H]
\centering
\epsfig{file=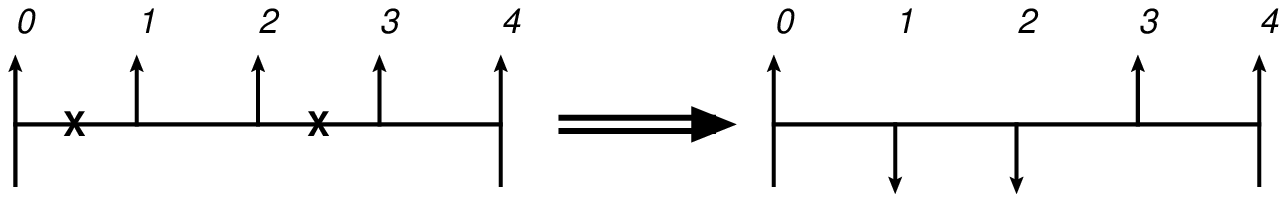}
\caption{Example of double twisted diagram with twists in the channels 1 and 3, which in the expansion Eq.\eqref{generalexapnsion} corresponds to the term $\tau_1\tau_3$.}
\label{doublytwisted}
\end{figure}

The meaning of the "twist" or "crossed line" is the same as before. By twist we mean that the diagram is rotated around the direction of the exchanged momenta to the right of the cross ("{\bf X}") sign. For example if one twists the diagram with respect to channel 1 (corresponding to $\omega_1$), the result is as shown in Fig. \ref{firsttwist}.
\noindent
We can generalize the twisting of the diagram in order to reach other channels. For example, in Fig.\ref{doublytwisted} we rotated twice. We move from left to right. The first twist brings the diagram similar to presented in Fig.\ref{firsttwist} and the second twist (cross in channel 3) rotates back the rest of the diagram to the right of the cross sign. It is important to stress that despite the fact that we rotate the diagram, it remains planar.
\noindent
The diagram in Fig.\ref{doublytwisted} corresponds to the following kinematic region:
\begin{eqnarray}
s_1 < 0, s_2 >0, s_3<0, s_4 >0; s_{012} < 0, s_{123} < 0, s_{234} < 0; s_{0123} > 0, s_{1234} < 0;s >0.\nonumber\\
\end{eqnarray}

%%%%%%%%%%%%%%%%%%%%%%%%%%%%%%%%%%%%%%%%%%%%%%%%%%%%%%%%%%%%%%%%%%%%%%%%%%%%%%%%%%%%%%%%%%%%%%%%5

\subsection{BDS predictions: examples}

Let us begin with a brief review of the five point and the six point functions in the multi-Regge kinematics. As shown in \cite{Bartels:2008ce}, for the $2\to3$ amplitude in the region of positive energies (no $\tau$ factors) we have the simple exponential form 
\begin{eqnarray}\label{BDSphase2to3}
\frac{M^{BDS}_{2\rightarrow 3}}{\Gamma(t_1) |s_1|^{\omega_1}|\Gamma_{1,2}| |s_2|^{\omega_2}\Gamma(t_2)}\;=\;e^{-i\pi\omega_1}e^{i\pi\omega_a}e^{-i\pi\omega_2}.
\end{eqnarray}
Analogous expressions hold for the other regions. The exponents resemble those which we have discussed in the previous section. However, in contrast to our discussion of the Regge pole framework, for the BDS amplitudes we can formulate simple rules which also fix the signs of the exponents of the production vertices. Let us next consider the $2\to4$ case in the region belonging to the coefficient $\tau_1\tau_3$ (Mandelstam region). From \cite{Bartels:2008ce,Lipatov:2010qf} we have
\begin{eqnarray}\label{BDSphase2to4}
\frac{M^{BDS}_{2\rightarrow 4}}{\Gamma(t_1) |s_1|^{\omega_1}|\Gamma_{1,2}||s_2|^{\omega_2}| |\Gamma_{2,3}||s_3|^{\omega_3} \Gamma(t_3)}\;=\;C e^{-i\pi\omega_2}e^{i\pi(\omega_a+\omega_b)}.
\end{eqnarray}
Here $C$ is the new phase factor, related to the one-loop approximation of the Regge cut
\begin{eqnarray}
C\;=\;e^{i\pi\left(\frac{\gamma_K}{4}\ln\frac{|q_1|^2|q_3|^2}{|k_a+k_b|^2\lambda^2}\right)}
\end{eqnarray}
with $k_a+k_b\;=\;q_1-q_3$. The remaining parts of the phases are obtained from the rules of Sec. II. It has been noticed in \cite{Lipatov:2010qf} that when combining this phase with the two vertex factors one arrives at a conformal invariant phase 
\be\label{delta_ph}
C e^{i\pi (\omega_a + \omega_b)} = e^{i \delta}
\ee
with
\begin{eqnarray}
\delta\;=\;\pi\frac{\gamma_K}{4}\ln\frac{|q_1||q_2||k_a||k_b|}{|k_a+k_b|^2|q_2|^2}.
\end{eqnarray}
It is important to recall the origin of the phase factor $C$: the BDS formula for the $2\to4$ amplitude  contains three $Li_2$ functions (dilogarithms) which depend upon the three independent anharmonic cross ratios. In the multi-Regge limit, one of these anharmonic cross ratios is a phase factor
\be
\Phi= \frac{(-s_2) (-s)}{(-s_{012})(-s_{123})}, 
\label{phase2to4}
\ee
with
\be
\Phi - 1 = \frac{|k_a+k_b|^2}{s_{2}},
\ee 
whereas the remaining two ratios go to zero. The dilogarithm depending upon the phase $\Phi$
appears in the combination
\begin{equation}
R(\Phi)\;=\;-\frac{1}{4}\ln^2 \Phi -\frac{1}{2}\ln \Phi \left(\ln \frac{(-t_{1})(-t_{3})}{(-s_{2})\mu^2}-\frac{1}{\epsilon}\right)-\frac{1}{2} Li_2(1-\Phi).
\end{equation} 
It is easy to see that 
\be
R(\Phi=1) = 0,
\ee
whereas for $\Phi=e^{\pm 2i\pi}$ the argument of the dilogarithm passes through a cut and    
\be
Li_2(1-\Phi) \to \mp 2\pi i \ln(1-\Phi)
\ee
with $\ln(1-\Phi)$ being real valued. Concluding, one can see that the analytical continuation of the combination of the  $Li_2$ function with the appropriate logarithms produces a logarithmic phase factor
\begin{eqnarray}
R(|\Phi|e^{\mp 2\pi i})\;=\;\pm i\pi\left(\ln\frac{|q_i|^2|q_j|^2}{ |q_i-q_j|^2 \lambda^2 }\right),
\label{pmV13}
\end{eqnarray}
which corresponds to the Mandelstam cut in the one-loop approximation.. There is an overall factor $\gamma_K/4$ in front of the logarithm, which was omitted during the computation of $R$ and should be restored in the final expression. 

For the $2\to5$ amplitude there are three phases which have to be rotated. We first consider the kinematic region belonging to the coefficient of $\tau_1\tau_3$. Here we rotate only 
\be
\Phi_1 =  \frac{(-s_{12}) (-s_{0123})}{(-s_{012})(-s_{123})}  
\ee
with
\be
\Phi_1 - 1 = \frac{|k_a+k_b|^2}{s_{12}},
\ee
whereas the two other phases are kept fixed. 
The BDS prediction is
\begin{eqnarray}\label{BDSphase2to5tau1tau3}
\frac{M^{BDS}_{2\rightarrow 5}}{\Gamma(t_1) |s_1|^{\omega_1}|\Gamma_{1,2}||s_2|^{\omega_2} |\Gamma_{2,3}| |s_3|^{\omega_3}|\Gamma_{3,4}| |s_4|^{\omega_4} \Gamma(t_4)}\;=\;C_{13}e^{-i\pi(\omega_2+\omega_4)}e^{i\pi(\omega_a+\omega_b+\omega_c)}
\end{eqnarray}
with 
\be
C_{13}\;=\;e^{i\pi\left(\frac{\gamma_K}{4}\ln\frac{|q_1|^2|q_3|^2}{|k_a+k_b|^2\lambda^2}\right)}\;\;\;;\;\;\;\;|k_a+k_b|^2\;=\;|q_1-q_3|^2.
\ee
We introduce the conformal invariant phase $\delta_{13}$:
 \be
C_{13} e^{-i\pi(\omega_2+\omega_4)}e^{i\pi(\omega_a+\omega_b+\omega_c)}= e^{-i\pi(\omega_2+\omega_4)}
e^{i\pi \omega_c} e^{i\delta_{13}} ,
\ee   
where
\be
\delta_{13} = \pi\frac{\gamma_K}{4}\ln\frac{|q_1||q_3||k_a||k_b|}{|k_a+k_b|^2|q_2|^2}.
\ee
The coefficient of $\tau_{2}\tau_4$ (with the rotating phase $\Phi_2$) is obtained from symmetry considerations. Next the region belonging to $\tau_1\tau_4$. The relevant phase which rotates to $e^{-2i\pi}$ is 
\be
\tilde{\Phi} =  \frac{(-s_{123}) (-s)}{(-s_{0123})(-s_{1234})}\;\;\;;\;\;\;\tilde{\Phi} - 1 = \frac{|k_a+k_b+k_c|^2}{s_{123}},
\ee
and the corresponding $Li_2$-function yields the phase factor
\be
C_{14}\;=\;e^{i\pi\left(\frac{\gamma_K}{4}\ln\frac{|q_1|^2|q_4|^2}{|k_a+k_b+k_c|^2\lambda^2}\right)}\;\;\;;\;\;\;\;|k_a+k_b+k_c|^2\;=\;|q_1-q_4|^2.
\ee
The prediction of the BDS formula for this kinematic region is
\begin{eqnarray}\label{BDSphase2to5tau1tau4}
\frac{M^{BDS}_{2\rightarrow 5}}{\Gamma(t_1) |s_1|^{\omega_1}|\Gamma_{1,2}||s_2|^{\omega_2} |\Gamma_{2,3}||s_3|^{\omega_3}  |\Gamma_{3,4}||s_4|^{\omega_2} \Gamma(t_4)}\;=\;C_{14} e^{-i\pi(\omega_2+\omega_3)}e^{i\pi(\omega_a+\omega_b+\omega_c)}.\nonumber\\
\end{eqnarray}
We write this as 
\be
C_{14} e^{-i\pi(\omega_2+\omega_3)}e^{i\pi(\omega_a+\omega_b+\omega_c)}= e^{-i\pi(\omega_2+\omega_3)}
e^{i\pi \omega_b} e^{i\delta_{14}} 
\ee
with the conformal invariant phase
\be
\delta_{14} = \pi\frac{\gamma_K}{4}\ln\frac{|q_1||q_4||k_a||k_c|}{|k_a+k_b+k_c|^2|q_2||q_3|}.
\ee
One can spot that the contribution for a {\it single} $Li_2$ function belonging to a Mandelstam cut is given by the simple exponential expression [cf.(\ref{pmV13})]
\begin{eqnarray}
C_{ij}\;=\;e^{i\pi\left(\frac{\gamma_K}{4}\ln\frac{|q_i|^2|q_j|^2}{|q_i-q_j|^2\lambda^2}\right)}.
\end{eqnarray}
The composite state of several single coefficients $C_{ij}$ consists of a product of $C$'s with appropriate signs of exponents, in accordance with the direction of the rotation of the analytical continuation.

Finally the coefficient of $\tau_1\tau_2\tau_2\tau_4$. Now we rotate $\tilde{\Phi}$ by $e^{-2i\pi}$ and 
$\Phi_1$ and $\Phi_2$ by $e^{+2i\pi}$. In terms of a single coefficient $C_{ij}$, the composite coefficient $C_{1234}$ will be:
\begin{eqnarray}
C_{1234}\;=\;C^{+}_{14}C^{-}_{13}C^{-}_{24},
\end{eqnarray}
where $C_{14}$ corresponds to the rotation of $\tilde{\Phi}$, $\Phi_1$, and $\Phi_2$ respectively. $\pm$ corresponds to the sign in front of $i\pi$ in the exponent.
We obtain
\begin{eqnarray}\label{BDSphase2to5tau1tau2tau3tau4}
\frac{M^{BDS}_{2\rightarrow 5}}{\Gamma(t_1) |s_1|^{\omega_1}|\Gamma_{1,2}||s_2|^{\omega_2} |\Gamma_{2,3}||s_3|^{\omega_3}|\Gamma_{3,4}||s_4|^{\omega_4} \Gamma(t_4)}\;=\;C_{1234} e^{-i\pi(\omega_a+\omega_b+\omega_c)}
\end{eqnarray}
with 
\be
C_{1234} = e^{-i\pi\left(\frac{\gamma_K}{4}\ln\frac{|q_2|^2|q_3|^2|k_a+k_b+k_c|^2}{|k_a+k_b|^2|k_b+k_c|^2\lambda^2}\right)}
\ee
and 
\be
C_{1234} e^{i\pi (\omega_a + \omega_c)} = e^{i\delta_{1234}}\;\;\;\text{with}\;\;\;\delta_{1234} = \pi\frac{\gamma_K}{4}\ln\frac{|q_1||q_4||k_a+k_b|^2|k_b+k_c|^2}{|k_a+k_b+k_c|^2|k_a||k_c||q_2||q_3|}.
\ee
In general, the definition of the phases $\delta_{ij...}$ is not unique. It depends upon which vertex factors are combined with the phases resulting from the $Li_2$ functions. We will fix these phases at the end of Sec. VC, after we have 
defined our renormalized Regge pole contributions.   

%%%%%%%%%%%%%%%%%%%%%%%%%%%%%%%%%%%%%%%%%%%%%%%%%%%%%%%%%%%%%%%%%%%%%%%%%%%%%%%%%%%%%%%%%%%%%%%%%%%%%%%%%%%%%%

\subsection{Propagators, Vertices, and $Li_2$ functions}

\begin{figure}[H]
\centering
\epsfig{file=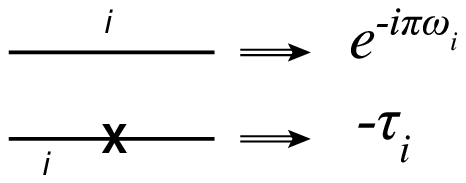}
\caption{Two types of propagators in channel $i$.}
\label{propagatorsBDS}
\end{figure}

\begin{figure}[H]
\centering
\epsfig{file=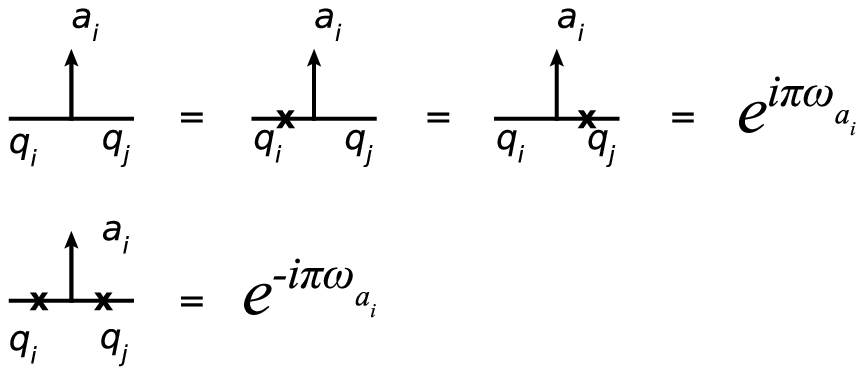}
\caption{Four types of vertices for the production of a particle with momentum $k_{a_{i}}$.}
\label{verticesrulesBDS}
\end{figure}

In order to generalize this discussion, we introduce "Feynman rules" for the calculation of the %coefficients of each monomial 
terms in the generating function. From the previous discussion it follows that there are three building blocks: propagators, vertices, and phases resulting from the $Li_2$ functions.
Beginning with the propagators, there are two types of propagators: one corresponds to untwisted $t$ channel lines, the other one to a twisted  line (Fig.\ref{propagatorsBDS}). For each untwisted propagator one should put $e^{-i\pi\omega_i}$, and for the twisted propagator, one puts $-\tau_i$.
\noindent
The second ingredient is the production vertex for the particle $a_i$ with the phase $\pi\omega_{a_i}$. We denote the produced momenta as $k_{a_1}$, $k_{a_2}$, $k_{a_3},\,...$. There are four types of vertices. Three vertices are {\it simple} - with at most only one twisted propagator line (upper line in Fig \ref{verticesrulesBDS}), and the rule is $e^{i\pi\omega_{a_i}}$. For the "doubly-twisted" vertex (the lower line in the Fig \ref{verticesrulesBDS}), we have the conjugated rule $e^{-i\pi\omega_{a_k}}$.
\noindent
For completeness we recapitulate the expressions for the different $\omega$'s presented here. The propagator in Fig.\ref{propagatorsBDS} corresponds to the Regge trajectory, which is given by
\begin{eqnarray}\label{regge_omega}
\omega_i\;=\;-\frac{\gamma_K}{4}\ln\frac{|q_i|^2}{\lambda^2},
\end{eqnarray}
while the vertex function $\omega_{a_i}$ corresponds to
\begin{eqnarray}\label{omega_p}
\omega_{a_i}\;=\;-\frac{\gamma_K}{8}\ln\left(\frac{|q_i|^2|q_j|^2}{|q_i-q_j|^2\lambda^2}\right)
\;\;;\;\;(j=i+1),
\end{eqnarray}
where $q_i-q_j\;=\;k_{a_i}$. 

\begin{figure}[H]
\centering
\epsfig{file=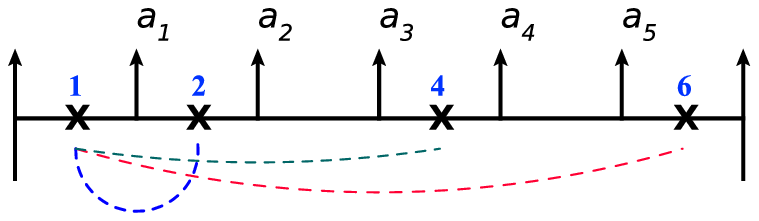}
\caption{Rules for obtaining the  $Li_2$ functions for a particular kinematic region (see text).}
\label{Li2example}
\end{figure}

The final ingredient is the phase resulting from the $Li_2$ functions. It depends on the kinematic regions, and it is convenient to find graphical rules for deriving these contribution. The idea of twisting the diagram is equivalent to changing the kinematic regions of energy variables $s_{ij..k}$. Consider the diagram in Fig.\ref{Li2example}.
\noindent
We connect crosses by lines. Each connecting line except for those which embrace a single production vertex  -   corresponds to a phase (anharmonic cross ratio) that has been rotated, $\Phi \to e^{\pm 2\pi i}$, and for each rotated phase the corresponding  $Li_2$ function has to be analytically continued and produces a nonvanishing phase. The sign in the exponent can be determined by counting the number of crosses embraced by the line: if the number is even, we have  $\Phi \to e^{ - 2\pi i}$ otherwise, $\Phi \to e^{+2\pi i}$. A simple example has already been given above, the case $2 \to4$, 
\begin{figure}[H]
\centering
\epsfig{file=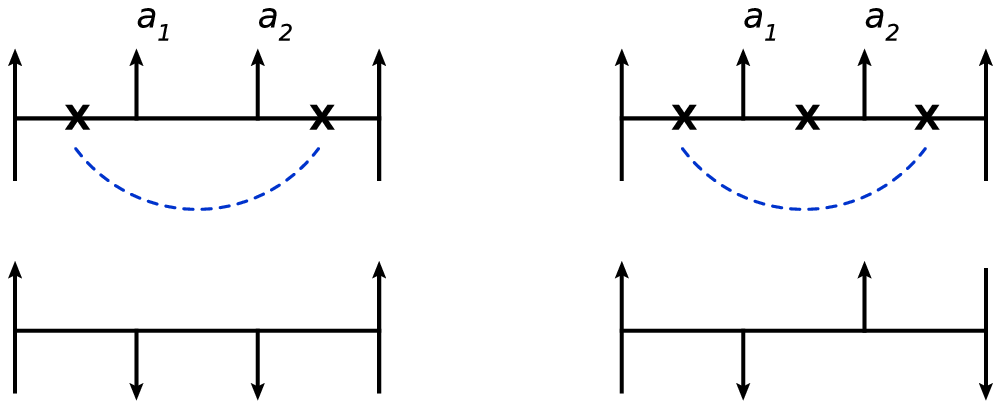}
\caption{Example for the relation between connecting lines and kinematic regions}
\label{6pt_Li2example}
\end{figure}
\noindent
For the coefficient $\tau_1 \tau_3$ (left part of Fig.\ref{6pt_Li2example}) 
there is only one such line which corresponds to the phase $\Phi=\frac{(-s)(-s_2)}{(-s_{012})(-s_{1223})}$ [Eq.\eqref{phase2to4}], and there is no cross ("zero cross") inside the line. This phase is rotated by $\Phi \to e^{ - 2\pi i}$. The analytic continuation of the $Li_2$ function leads to the expression Eq.\eqref{pmV13} which we denote by the "potential" $V_{13}$,
\be
e^{i\pi V_{13}}=e^{i\pi \frac{\gamma_K}{4} \ln \frac{|q_1|^2 |q_3|^2}{|q_1-q_3|^2\lambda^2}}.
\ee
If we apply the same discussion to the coefficient of $\tau_1\tau_2\tau_3$ (right part of Fig.\ref{6pt_Li2example}), 
we have one cross inside the line, the phase is rotated by $\Phi \to e^{ +2\pi i}$, and the  analytic continuation of the $Li_2$ function gives
\be
e^{-i\pi V_{13}}=e^{-i\pi \frac{\gamma_K}{4} \ln \frac{|q_1|^2 |q_3|^2}{|q_1-q_3|^2\lambda^2}}
\ee
We generalize the notion of a "potential" for the interaction between two crosses in the $t_i$ channel and the $t_j$ channel:
\begin{eqnarray}
\label{vij}
V_{ij}\;=\;\frac{\gamma_K}{4}\ln\frac{|q_i|^2|q_j|^2}{|q_i-q_j|^2\lambda^2}.
\end{eqnarray}
Returning to the production vertices $\omega_{a_i}$, it is convenient to extend the notion of the "potential" also to 
neighboring lines which encircle not more than one production vertex:  
\ba
V_{i i+1}= - 2 \omega_{a_{i}}.
\ea
With this definition we modify our rules for the production vertex: instead of writing $e^{\pm i \pi \omega_{a_i}}$  (depending on whether we have crosses on both sides of the produced particle $a_i$) we adopt the following rule:
for each vertex we write the unique factor $e^{+i \pi \omega_a}$, and for production vertices with crosses on both sides we include the 
additional factor 
\be
e^{i\pi V_{ii+1}}.
\ee
This allows us to include into our rules, in  Fig.\ref{Li2example}, also the short line around the vertex $a_1$: 
now each line that connects crosses in the $t_i$, and the $t_j$ channel obtains a factor
\be
e^{\pm i \pi V_{ij}}.
\ee 
If the channels $i$ and $j$ are adjacent (i.e. $j=i+1$ and they enclose a production vertex) the sign is always 
positive. Otherwise the counting rules of crosses inside the lines apply (Fig.\ref{Li2_ph}),  
\begin{figure}[H]
\centering
\epsfig{file=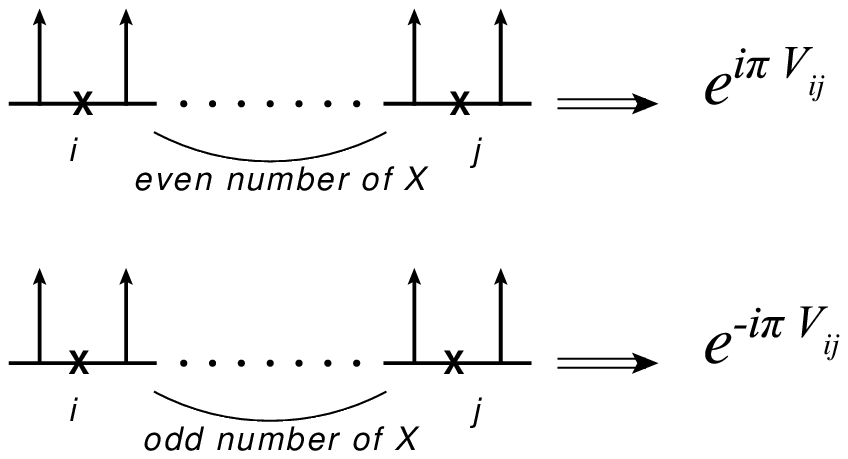}
\caption{Sign of the phase depending on the number of crosses "{\bf X}" between two twists $i$ and $j$.}
\label{Li2_ph}
\end{figure}
\noindent
Concluding everything, we formulate Feynman-like rules for the calculation of the coefficients in the monomial expansion of the amplitude $A_{BDS}$ [Eq.\eqref{generalexapnsion}]. Each coefficient $a_{i..j}$ will be written 
in the form 
\be
a_{i..j}=\pm |a_{i..j}| e^{i\varphi_{i...j}} e^{i\delta_{i...j}},
\ee
and for the overall sign and for the sum of the phases in the exponent  we have the following rules:
\begin{itemize}
\item for each $t$ channel we write a propagator (twisted and untwisted) according to the rules
\begin{itemize}
\item twisted propagator: $\rightarrow$ $-1$
\item untwisted propagator in channel $t_i$: $\rightarrow$ $e^{-i\pi\omega_i}$
\end{itemize}
\item write the product of phase factors of vertices for all produced particles:
 $e^{i\pi\left(\omega_{a_1} + \omega_{a_2} +...\right)}$
\item write all pairwise interactions $e^{\pm i\pi V_{ij}}$, $i\neq j$ with the sign $(-1)^n$ in the exponent.
Here $n$ is the number of crosses encircled by the pair $(ij)$.
\end{itemize}
These rules uniquely define the sum of all phases. For our purposes, however, we go one step further and divide this sum into two terms, $i(\varphi_{i...j} + \delta_{i...j})$. Examples have been given in Sec. IIIB for the case $2\to4$. The first part, $i\varphi_{i...j}$, contains all the propagators, and it may contain some of the production vertices. The second part has to be conformal invariant. From these requirements alone, we do not find a unique separation into the two terms, $i(\varphi_{i...j} + \delta_{i...j})$. We will come back to this question  in our final Sec. IVD. As an example of applying these rules, we return to the diagram in Fig.\ref{Li2example}:
\begin{itemize}
\item propagators: $(-)(-)e^{-i\pi\omega_3}(-)e^{-i\pi\omega_5}(-)$
\item vertices: $ e^{i\pi\left(\omega_{a_1}+\omega_{a_2}+\omega_{a_3}+\omega_{a_4}+\omega_{a_5}\right)}$
\item potentials:  $e^{i\pi\left(V_{12} - V_{14} + V_{16} + V_{24} - V_{26} + V_{46}\right)} $.
\end{itemize}
The final expression for Fig.\ref{Li2example} becomes
\begin{eqnarray}\label{rules4twists}
e^{i\pi\left(\omega_{a_1}+\omega_{a_2}+\omega_{a_3}+\omega_{a_4}+\omega_{a_5}\right)}e^{-i\pi\left(\omega_3+\omega_5\right)}e^{i\pi
\left(V_{12} - V_{14} + V_{16} + V_{24} - V_{26} + V_{46}\right)}\tau_1\tau_2\tau_4\tau_6.
\end{eqnarray}
The logarithmic form of the potential Eq.\eqref{vij}, together with the exponential form of the 
coefficient of the monomial in Eq.\eqref{rules4twists}, allow an interesting analogy.
Namely, we can interpret $V_{ij}$ as a two dimensional Coulomb potential of the interaction of two point charges $i$ and $j$, derived from the Polyakov string action.
In more detail, we consider the product of $k$ vertex operators, i.e. correlators of the form 
\begin{eqnarray}
\langle 0|e^{i\pi\sum_{r=1}^{k} c_r \left[\phi(\vec{\rho}_{r})-\phi(\vec{\rho}_{0}) \right]}|0\rangle,
\label{correlator}
\end{eqnarray}
where the averaging is done with the free action
\begin{eqnarray}
e^{i\frac{1}{2}\int\, d^{2}\vec{\rho} \left[\partial_{\sigma}\phi(\vec{\rho})\right]^2}
\end{eqnarray}
and $c_r\;=\;(-1)^r$ is the charge. It is convenient to introduce the following currents:
\begin{eqnarray}
\pi\sum_{r=1}^{k}c_r\left[\phi(\vec{\rho}_{r})-\phi(\vec{\rho}_{0})\right]\;&=&\;\pi\int d^2\vec{\rho}\,\phi(\vec{\rho}_{r})\sum_{r=1}^{k}c_r\left(\delta^{(2)}(\vec{\rho}-\vec{\rho}_r)-\delta^{(2)}(\vec{\rho}-\vec{\rho}_0)\right)\;=\;\nonumber\\ &=&\int d^2\vec{\rho}\,\phi(\vec{\rho})J(\vec{\rho}),
\end{eqnarray}
with
\begin{eqnarray}
J(\vec{\rho})\;=\;\pi\sum_{r=1}^{k}c_r\left(\delta^{(2)}(\vec{\rho}-\vec{\rho}_r)-\delta^{(2)}(\vec{\rho}-\vec{\rho}_0)\right).
\end{eqnarray}
One can calculate the Gaussian integral of the neutral system 
\begin{eqnarray}
Z[J]= \int e^{i\int d^2\vec{\rho}\left[\frac{1}{2} \left(\partial_{\sigma}\phi(\vec{\rho}\right)^2 + \phi(\vec{\rho})J(\vec{\rho})\right]}\mathcal{D}\phi
\end{eqnarray}
by using the inverse of the two-dimensional Laplacian,
\begin{eqnarray}
\partial^{2}_{\sigma}\tilde{\phi}(\vec{\rho})\;=\;J(\vec{\rho})\;\rightarrow\;\;\tilde{\phi}(\vec{\rho})=\frac{1}{4\pi}\int d^2\vec{\rho'} J(\rho')\log\left(|\vec{\rho}-\vec{\rho'}|^2\right)
\end{eqnarray}
and by shifting the field variables: $\phi=\phi'+\tilde{\phi}$. One obtains:
\begin{eqnarray}
\int d^2\vec{\rho}\left[\frac{1}{2}\left(\partial_{\sigma}\phi(\vec{\rho})\right)^2+\phi(\vec{\rho}) J(\vec{\rho}) \right]\;&=&
\nonumber\\
\;=\frac{1}{8\pi}\int\int \ d^2\vec{\rho}d^2\vec{\rho'}J(\vec{\rho})\log\left(|\vec{\rho}-\vec{\rho'}|^2\right)J(\vec{\rho'})&+&\frac{1}{2}\int d^2\vec{\rho}\left(\partial_{\sigma}\phi'\right)^2.
\end{eqnarray}
From this expression one derives, for the correlator (\ref{correlator}), an exponential of the form:
\begin{eqnarray}
V_{ij}\;=\;\frac{\pi}{8}\sum_{r,r'=1}^{k}c_r\,c_{r'}\left[\log|\vec{\rho_r}-\vec{\rho_r'}|^2-\log|\vec{\rho_r}-\vec{\rho_0}|^2-\log|\vec{\rho_{r'}}-\vec{\rho_0}|^2+\log|\vec{\rho_{00}}|^2\right].
\end{eqnarray}
In the first term one recognizes the logarithmic part of the "potential" $V_{ij}$ between two crosses defined in (\ref{vij}). In particular, we notice the universal short-range interaction between two adjacent crosses,
\begin{eqnarray}
V_{i,i+1} \sim \log|\vec{\rho_{i}}-\vec{\rho}_{r_{i+1}}|^2.
\end{eqnarray}
Finally, returning to the generating function introduced at the beginning of this section,
\ba
A_{BDS}\;=\;a_0 + a_1\tau_1 + a_2\tau_2 +...+a_n\tau_n + a_{12}\tau_1\tau_2 + a_{13}\tau_1\tau_3+...a_{1...n}\tau_1\tau_2...\tau_n,
\nonumber
\ea
we can interpret this expression also as a partition function, where each terms represents one of the correlators described above.  For the rest of this paper, we will not pursue this analogy any further.

\section{Subtractions from Regge pole contributions}

In the previous sections we have seen that the Regge pole formula, based upon 
factorization and the analytic decomposition into 5 terms (for the case $2\to4$) or 14 terms  (for the case $2\to 5$), exhibits, when continued into different kinematic regions with positive and negative energies, terms with unphysical singularities. At the end of Sec. III we have indicated that Regge cut terms are needed in order to compensate these unwanted singularities. The subsequent analysis of the BDS predictions, on the other hand, has shown that the BDS formula is not in agreement with the Regge pole structure, because it contains contributions from the $Li_2$ functions.
As a consequence, depending on the kinematic region, it contains  phases which, in the $2\to4$ case \cite{Bartels:2008ce}, have been understood as a signal of the beginning of Regge cut contributions. 
In this final section we concentrate on the case $n=7$, and we show that Regge cut contributions 
can be determined which satisfy the following two conditions: 
\begin{enumerate}
\item the terms with Regge cuts have the correct phase structure for absorbing the unwanted pole terms,
\item after absorbing the unphysical pole pieces of the Regge poles into the Regge cut terms,
we are left with conformal invariant Regge pole contributions.
\end{enumerate}
\noindent
To be definite, our construction proceeds as follows. Initially we have the Regge pole terms which, as we have 
stated, factorize in the kinematic region of positive energies but, when analytically continued, lead to unphysical singularities. They have to be absorbed into Regge cut contributions. Schematically we therefore write,
\be
A= A_{pole} + A_{cut},
\ee
where the pole contributions are listed in Appendix A, and the phase structure of the cut contributions have to be discussed in the following.
Their contributions to the scattering amplitude depend upon the kinematic region: they vanish for positive energies (and in the Euclidean region), and they are nonzero in exactly those kinematic regions where the Regge poles exhibit the unphysical 
singularities. After having fixed the subtractions we will arrive at modified expressions,
\be
A= A'_{pole} + A'_{cut},
\ee   
where the primes indicate that, in each physical region with Regge cuts and singular Regge pole pieces, the unphysical singular pieces have been absorbed by the Regge cuts. In this new representation the amplitude, for each region $\tau_i...\tau_j$,  will be written in the form:
\be
A= A_{BDS} R, 
\ee 
where  $A_{BDS}$ contains the phase factors $\varphi_{i...j}$ and $\delta_{i...j}$ calculated in the previous section (IIIC), 
and the conformal invariant remainder function $R$ is of the form  
\be
R e^{i\delta} = {\text{conformal pole + conformal Regge cut}}. 
\ee
For illustration we return, once more,  very briefly to the $2\to4$ case \cite{Lipatov:2010qf}. As shown in  \cite{Bartels:2008ce,Bartels:2008sc}, the Regge cut piece has the phase $i e^{-i\pi \omega_2}$. 
To see this we remind that, in the decompositon  of the $2\to4$, amplitude, the Regge cut appears in two 
of the five terms. Their phase structure follows from the energy factors which, in the notation of  \cite{Bartels:2008ce,Bartels:2008sc}, is
\ba
W_3 \sim (-s_2)^{\omega_{21}} (-s_{012})^{\omega_{13}} (-s)^{\omega_3} V_{cut}\nonumber \\
W_4  \sim     - (-s_2)^{\omega_{23}}  (-s_{123})^{\omega_{31}} (-s)^{\omega_1} V_{cut}.
\ea
The coefficient $V_{cut}$ is the same in both terms, and there is relative minus sign between the two 
partial waves. From this structure one derives easily that the sum of these two contributions vanishes in the physical region where all energies are positive (a phase factor $e^{-i\pi}$ form each energy), in the Euclidean region (all energies negative, i.e. all phases reduce to unity), and also in the region where only one energy is negative.  In contrast to this, in the region $s, s_2>0$, $s_{012}, s_{123}<0$, the sum is proportional to $i e^{-i\pi \omega_2}$. 
On the other hand, the Regge pole, when continued into this kinematic region, takes the form Eq.\eqref{2to4twist13a}, i.e. we have one term proportional to $e^{-i\pi \omega_2}$, and two terms proportional to $i e^{-i\pi \omega_2}$. The latter ones have the same phase structure as the Regge cut contribution, and thus they can be combined with the Regge cut: we can remove them by a special contribution ("subtraction") inside the Regge cut. What is then left is the first term of the Regge pole contribution
\be
e^{-i\pi \omega_2} \cos (\pi \omega_{ab}),
\ee
with $\omega_{ab}\;=\;\omega_a-\omega_b$. Here the argument of the "cosine" function is conformally invariant. Therefore, this expression defines, for this kinematic region, a  "conformal" Regge pole contribution. The amplitude can be written as 
\be
A=A_{BDS} R,
\ee
where $A_{BDS}$ contains the phase factor $e^{-i\pi \omega_2}$, and 
\be
R e^{i\delta}  = \cos (\pi \omega_{ab}) + i {\text{  ReggeCut}} .
\ee
The new Regge cut contribution is expected to be conformally invariant. 
%%%%%%%%%%%%%%%%%%%%%%%%%%%%%%%%%%%%%%%%%%%%%%%%%%%%%%%%%%%%%%%%%%%%%%%%%%%%%%%%%%%%%%%%%%%%%%%%%%%%%%%%%

\subsection{Analytic structure of the $2\to5$ production amplitude}
In the following we will extend this analysis to the $2\to5$ case. We now have three different Regge cut contributions.
They are illustrated in the following figure: 
\begin{figure}[H]
\centering
\epsfig{file=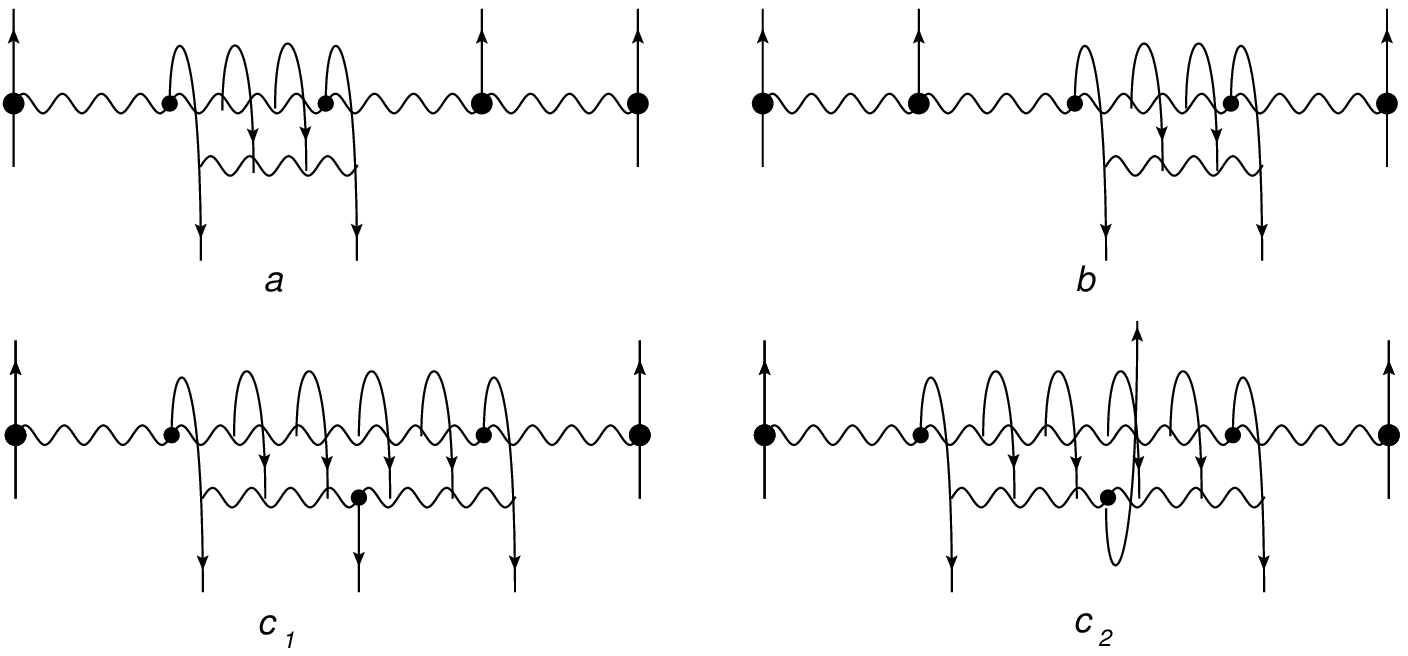}
\caption{Regge cut contributions for the $2\to5$ production amplitude}
\label{2to5Reggecuts}
\end{figure}
\noindent
In addition to the $t$ channels where the Regge cuts appear, we have also indicated a few kinematic regions in which these Regge cut contribute.  In the generating functional, these kinematic regions correspond to the coefficients of $\tau_1\tau_3$, $\tau_2\tau_4$, $\tau_1\tau_4$, and $\tau_1\tau_2\tau_3\tau_4$. The analytic representation of the $2\to5$ amplitude contains 14 different terms. 
They are illustrated below in Fig. \ref{f3_to_w3},  
\begin{figure}[H]
\centering
\epsfig{file=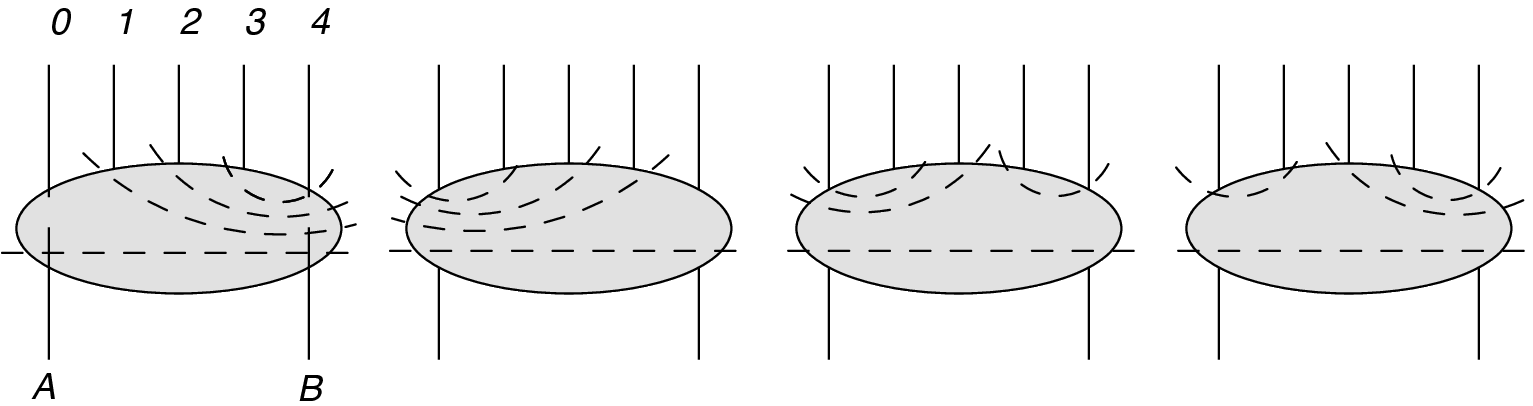}
\caption{Terms without Regge cuts}
\label{f3_to_w3}
\end{figure}

\begin{figure}[H]
\centering
\epsfig{file=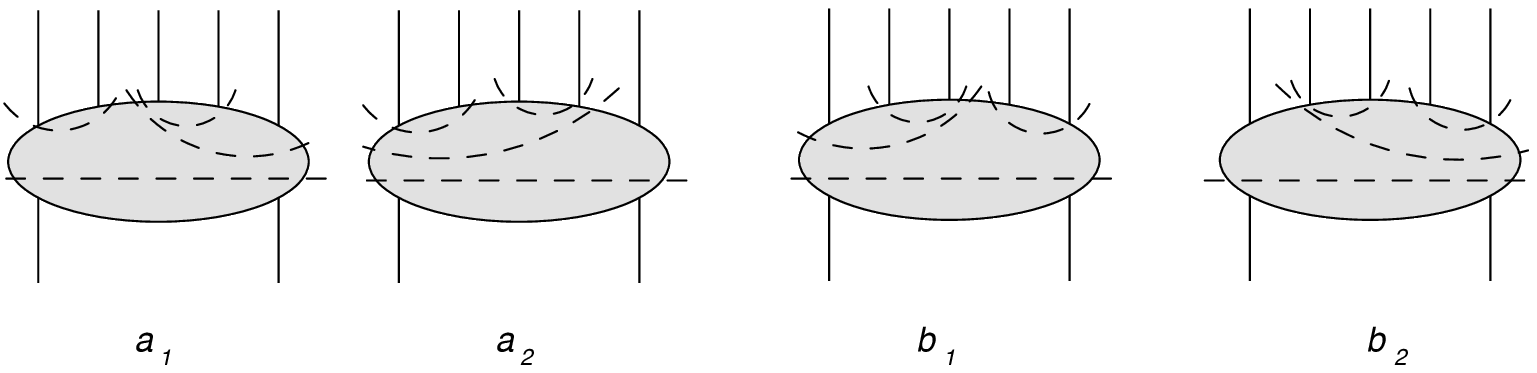}\\
\epsfig{file=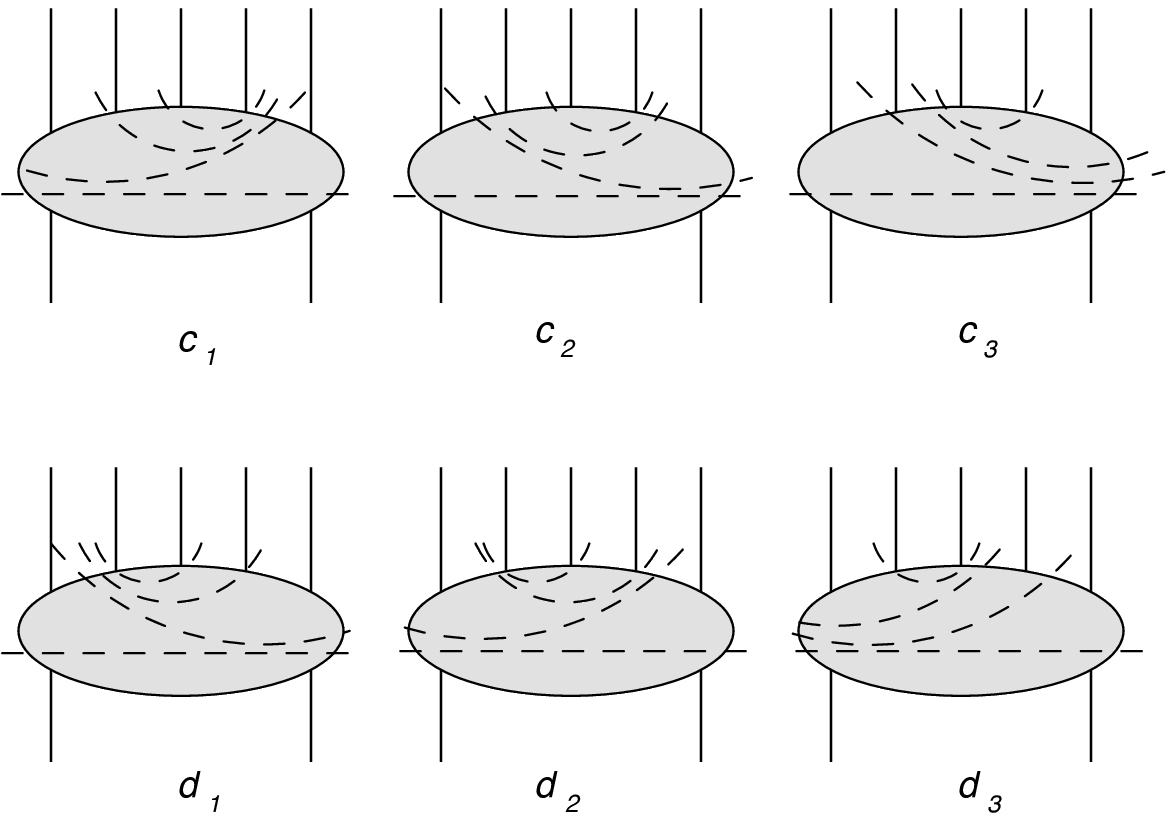}
\caption{Terms which contain Regge cut contributions: two doublets (a) and (b), and  two triplets (c) and (d)}
\label{f3_to_w3all3}
\end{figure}
\noindent
Here each term is written as a multiple Sommerfeld-Watson integral over $\omega$ variables, and the integrand comes as a product of energy factors which contain all the phases and a real-valued partial wave. For simplicity, we will disregard the $\omega$ integration in the rest of our paper.. 
The analytic structure of these terms is in agreement with the Steinmann relations, i.e. each of these 14 terms has a maximal set of energy discontinuities in nonoverlapping channels (denoted by dashed lines).  

Only 10 of these 14 terms contain Regge cut contributions: they can be arranged as two doublets $a,b$ and two triplets $c,d$. The "short" Regge cut in the $t_3$ channel [Fig.\ref{2to5Reggecuts}b] is contained in the first  doublet $a_1$ and $a_2$ and in the first triplet, $c_1$, $c_2$, and $c_3$. Similarly, the  "short" Regge cut in the $t_2$ channel [Fig.\ref{2to5Reggecuts}a]
is contained in the second doublet, $b_1$ and $b_2$, and in the second triplet, $d_1$, $d_2$, and $d_3$. Finally, the "long" cut in Fig.\ref{2to5Reggecuts}$c_1$, and Fig.\ref{2to5Reggecuts}$c_2$ appears in the first two terms of both triplets. In each term, these Regge cut contributions are additive. As an example,  the first two terms of the triplets are sums of two terms, each of a "short" cut and of the "long" cut.  

Next we are interested in the phase structure of these terms: it follows from the energy 
factors which we list in the following. For the doublets we have   
\begin{eqnarray}
\label{acuts}
a_1\;&=&\;\left(-s_{1}\right)^{\omega_{12}} \left(-s_{3}\right)^{\omega_{34}}\left(-s_{234}\right)^{\omega_{42}}\left(-s\right)^{\omega_2}\\ 
a_2\;&=&\;\left(-s_{1}\right)^{\omega_{12}} \left(-s_{3}\right)^{\omega_{32}}\left(-s_{0123}\right)^{\omega_{24}}\left(-s\right)^{\omega_4} \nonumber
\end{eqnarray}
and
\begin{eqnarray}
b_1\;&=&\;\left(-s_{2}\right)^{\omega_{21}} \left(-s_{012}\right)^{\omega_{13}}\left(-s_{4}\right)^{\omega_{43}}\left(-s\right)^{\omega_3}\\ 
b_2\;&=&\;\left(-s_{2}\right)^{\omega_{23}} \left(-s_{4}\right)^{\omega_{43}}\left(-s_{1234}\right)^{\omega_{31}}\left(-s\right)^{\omega_1}.\nonumber
\end{eqnarray}
Similarly for the two triplets,
\begin{eqnarray}
\label{ccuts}
c_1\;&=&\;\left(-s_{3}\right)^{\omega_{32}} \left(-s_{123}\right)^{\omega_{21}}\left(-s_{0123}\right)^{\omega_{14}}\left(-s\right)^{\omega_4}\\ \nonumber
c_2\;&=&\;\left(-s_{3}\right)^{\omega_{32}} \left(-s_{123}\right)^{\omega_{24}}\left(-s_{1234}\right)^{\omega_{41}}\left(-s\right)^{\omega_1}\\ \nonumber
c_3\;&=&\;\left(-s_{3}\right)^{\omega_{34}} \left(-s_{234}\right)^{\omega_{42}}\left(-s_{1234}\right)^{\omega_{21}}\left(-s\right)^{\omega_1} \nonumber
\end{eqnarray}
and 
\begin{eqnarray}\label{dcuts}
d_1\;&=&\;\left(-s_{2}\right)^{\omega_{23}} \left(-s_{123}\right)^{\omega_{34}}\left(-s_{1234}\right)^{\omega_{41}}\left(-s\right)^{\omega_1}\\ \nonumber
d_2\;&=&\;\left(-s_{2}\right)^{\omega_{23}} \left(-s_{123}\right)^{\omega_{31}}\left(-s_{0123}\right)^{\omega_{14}}\left(-s\right)^{\omega_4}\\ \nonumber
d_3\;&=&\;\left(-s_{2}\right)^{\omega_{21}} \left(-s_{012}\right)^{\omega_{13}}\left(-s_{0123}\right)^{\omega_{34}}\left(-s\right)^{\omega_4}. \nonumber
\end{eqnarray}
It should be noted that in these expressions, for simplicity, we have disregarded $\kappa$ factors as well as energy scales.
As an example,  the complete form of $d_1$ from Eq.\eqref{dcuts} which includes these coefficients has the 
form \cite{Bartels:2008ce,Bartels:2008sc}
\begin{eqnarray}
d_1\;&=&\;\left(\frac{-s_{2}}{\mu^2}\right)^{\omega_{23}} \left(\frac{-s_{123}\kappa_{23}}{\mu^4}\right)^{\omega_{34}}\left(\frac{-s_{1234}\kappa_{23}\kappa_{34}}{\mu^6}\right)^{\omega_{41}}\left(\frac{-s\kappa_{12}\kappa_{23}\kappa_{34}}{\mu^8}\right)^{\omega_1},\nonumber
\label{kappafactors}
\end{eqnarray}
where 
\begin{eqnarray}
\kappa_{ii+1}\;=\frac{s_i s_{i+1}}{s_{i-1 i i+1}}\;=\;|q_i - q_{i+1}|^2\;\;\;,\;\;\;\;\text{and the usual  convention:}\;\;\;s_i\equiv s_{i-1 i}.
\end{eqnarray}
As a result of these $\kappa$ factors, all energy factors $d_1$, etc. can be written in the 
common form 
\ba
d_1 = {\text{phase factor}}\times \left( \frac{|s_1|}{\mu^2}\right)^{\omega_1}
\left( \frac{|s_2|}{\mu^2}\right)^{\omega_2}\left(\frac{|s_3|}{\mu^2}\right)^{\omega_3}\left(\frac{|s_4|}{\mu^2}\right)^{\omega_4}
\ea
In the following, our interest will first be devoted to the phase factors derived from Eqs.\eqref{acuts} - \eqref{dcuts}: they depend upon the kinematic regions. In the next step, we will determine the coefficients that accompany the phase factors; they are real valued and do not depend upon the kinematic region we are considering.   

%%%%%%%%%%%%%%%%%%%%%%%%%%%%%%%%%%%%%%%%%%%%%%%%%%%%%%%%%%%%%%%%%%%%%%%%%%%%%%%%%%%%%%%%%%%%%%%%%%%%%%%%%5

\subsection{Determination of the coefficients of the partial waves}

As we have said before,the kinematic regions in which the Regge pole expressions 
(listed in Appendix A)  contain poles of the type $1/\sin (\pi \omega_i)$ are the same regions for which we also have Regge cut contributions \footnote{Conditions for the existence of the Regge cuts have been formulated in the appendix of \cite{Lipatov:2009nt}.}. For each such region we write schematically
\be
f = f_{pole} + f_{cut}.
\ee    
In this notation, $f$ denotes the sum of all those terms which contribute to this region, and it contains both the energy (and phase) factors and their real-valued coefficient, the partial waves.  As a consequence, the form of the $f$ will be different in different kinematic regions. In general, the Regge cut piece will a sum of several terms: for example, the coefficient of $\tau_1\tau_2\tau_3\tau_4$ 
contains the two "short"  cuts and the "long" cut:
\be
f_{cut} = f_{\omega_2} + f_{\omega_3} + f_{\omega_2 \omega_3}.
\ee
In this paper we will not address the full structure of these Regge cut terms. Instead, we will concentrate on the overall phases, $f^{phase}_{cut}$, and only those pieces of the Regge cuts which absorb the  "unwanted" pieces of the Regge pole contributions, i.e. those terms that have the unphysical poles of the form $1/\sin(\pi \omega_i)$: $\delta f_{cut}$, namely:
\begin{eqnarray}
\label{f_omega}
f_{\omega_i}\;=\;N_{\omega_i}f^{phase}_{\omega_i}\delta f_{\omega_i}\,.
\end{eqnarray}
We therefore have to keep in mind that the $f_{cut}$ which we discuss in the following  contain only the subtraction terms but not the full Regge cut terms. We will name this procedure "subtraction", in analogy to the removal of ultraviolet divergences in the renormalization of quantum field theory.   

In more detail, for the two doublets and for the two triplets, we will find a set of coefficients  which should satisfy the following requirements:\\
(i) the Regge cuts contribute only in specific kinematic regions where the so-called Mandelstam conditions are fulfilled.  In particular, Regge cuts do not contribute to the Euclidean region or to the physical region where all energies are positive.\\
(ii) Phases of the Regge cut contributions have to match the "unwanted" pieces of the Regge pole contributions, i.e. those terms which have the unphysical poles of the form $1/\sin(\pi \omega_i)$.\\
(iii) After having absorbed these "unwanted" pole terms into the Regge cut terms, the remaining Regge pole contributions have to be conformal invariant.       
  
Let us begin with the "short" Regge cut in the $t_3$ which appears in the terms labeled by $a_1$, $a_2$, $c_1$, $c_2$, and $c_3$. We are searching for real-valued coefficients of theses terms which, for the sum of all five terms, lead to correct phases in all kinematic regions. First we notice that, in the region of all energies being positive, 
all $c_i$ have the common phase $e^{-i\pi \omega_3}$,  and all $a_i$ the common phase $e^{-i\pi (\omega_1-\omega_2+\omega_3)}$. The absence of the Regge cut in this region implies that   
the sum of  the terms $a_1$, $a_2$ and the sum of the terms  $c_1$, $c_2$, $c_3$ must be 
zero separately. This alone does not fix the coefficients of the $c_i$. We make the ansatz (which will be justified in a moment) and choose, for the coefficients of $c_1$, $c_2$, and $c_3$, the relative weights $\frac{1}{2}$,  $\frac{1}{2}$, and $-1$, respectively. Similarly, for 
the coefficients of $a_1$ and $a_2$ the relative weights are $+1$ and $-1$, respectively. 
In order to determine the common factors of the $c_i$, we go to the region $\tau_2\tau_4$: here the terms $c_1$ and $c_2$ have the common phase $e^{-i\pi \omega_3} e^{-i\pi (\omega_4 -\omega_2)}$, whereas $c_3$ has the phase   $e^{-i\pi \omega_3} e^{-i\pi (\omega_2 -\omega_4)}$. Taking into account the relative weights given above, the sum of the terms $c_i$ gives the phase $e^{-i\pi \omega_3}  2i \sin \pi (\omega_2-\omega_4)$. In the same way, the sum of $a_1$ and $a_2$ lead to the factor $e^{-i\pi (\omega_3 + \omega_1-\omega_2}  2i \sin \pi (\omega_4-\omega_2)$. Combining the sum of the $c_i$ terms with the sum of the $a_i$ terms we still have the freedom to chose coefficients:
with the choice $\sin(\pi \omega_{2a})$ and $\sin(\pi \omega_{1a})$\footnote{Please keep in mind that $\omega_{ij}\;=\;\omega_i-\omega_j$.} we have, again for the region $\tau_2\tau_4$, the result: 
\ba
\sin(\pi \omega_{2a})\left\{\frac{1}{2} c_1 + \frac{1}{2} c_2 - c_3 \right\} + \sin(\pi \omega_{1a})\left\{a_1 - a_2\right\}
\nonumber\\
=2i \sin (\pi \omega_{12})  \sin(\pi\omega_{34})  e^{-i\pi \omega_1} e^{i\pi \omega_a} e^{-i\pi \omega_3}.
\ea
The phases are in agreement with what one expects from Regge factorization: the Regge cut in the $t_3$ channel has the same phase in the $2\to4$ amplitude, $i e^{-i\pi \omega_3}$, and the phase of the $t_1$ channel together with the production vertex of particle $a$ factorizes as $e^{-i\pi \omega_1} e^{i\pi \omega_a}$. 
    
\begin{figure}[H]
\centering
\epsfig{file=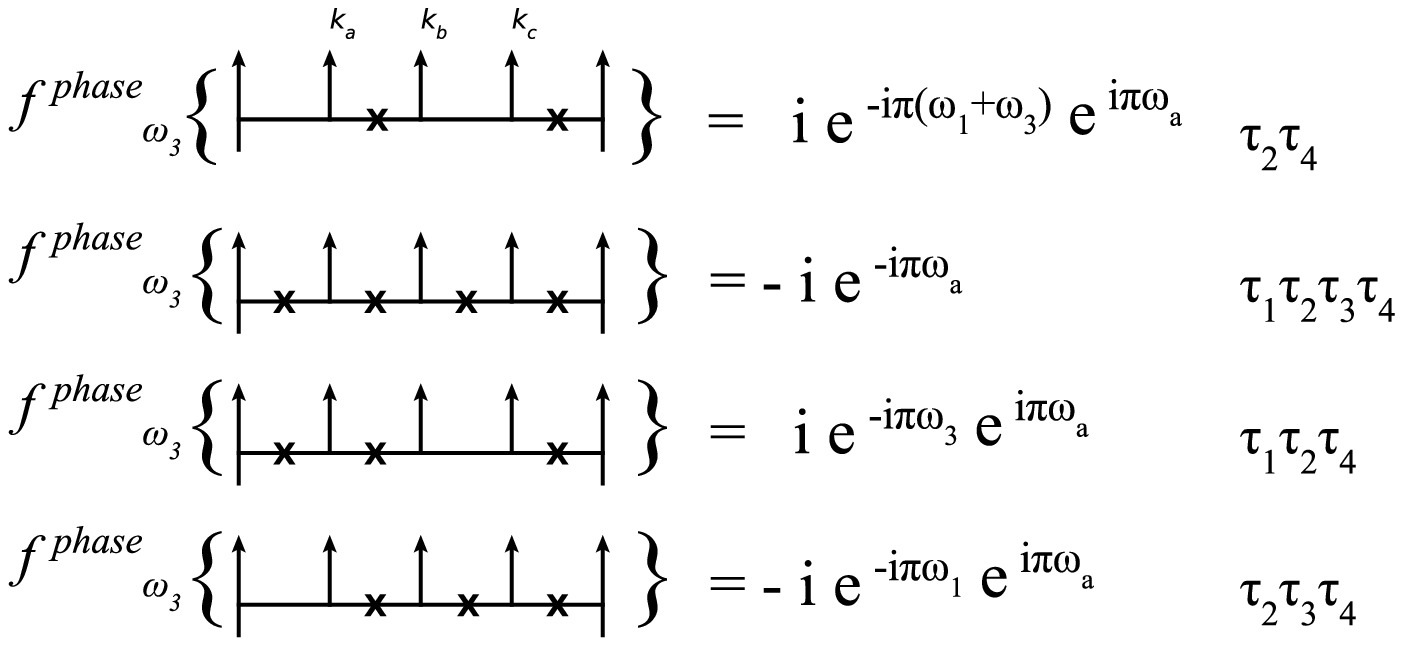}
\caption{Analytical continuation of $f_{\omega_3}$.}
\label{f3_cont}
\end{figure}

However, this is not yet the final answer for the cut in the $\omega_3$ channel. Namely, when going to the 
region $\tau_1\tau_4$,  we find the phases
\be
a_1- a_2 = e^{-i\pi \omega_3} 2i \sin(\pi \omega_{24})  
\ee
\be
\frac{1}{2}(c_1+c_2)-c_3=e^{-i\pi \omega_3} i \sin(\pi\omega_{14}).
\ee
Together with the prefactors $\sin (\pi \omega_{1a})$, $\sin (\pi \omega_{2a})$, these terms cannot be combined to arrive at the the expected phase $e^{-i\pi (\omega_2+\omega_3)}$. As a solution, we chose to completely cancel this contribution by adding a term proportional to $c_1-c_2$. In the region $\tau_1\tau_4$ we have
\be
c_1-c_2 =e^{-i\pi \omega_3} 2 i \sin(\pi \omega_{14}),
\label{c1-c2}
\ee 
and with the following coefficients we arrive at our final answer for the "short" cut in the $\omega_3$ channel, 
\begin{eqnarray}
\label{f_3phases}
N_{\omega_3} f_{\omega_3}^{phase} \;&=&\;\sin(\pi \omega_{2a})\left\{\frac{1}{2} c_1 + \frac{1}{2} c_2 - c_3 \right\} + \sin(\pi \omega_{1a})\left\{a_1 - a_2\right\}- \\ 
&-& \frac{1}{\sin(\pi \omega_{14})}\left(\frac{1}{2}\sin(\pi \omega_{14})\sin(\pi \omega_{2a}) + \sin(\pi \omega_{42})\sin(\pi \omega_{1a}) \right)\left\{c_1 - c_2\right\}. \nonumber
\end{eqnarray}
We make sure that, by analytically continuing this function  $f_{\omega_3}$ into different kinematic regions, we find correct answers. In detail, the results are the following: nonzero values appear only in the four kinematic regions $\tau_2\tau_4$, $\tau_1\tau_2\tau_3\tau_4$,
$\tau_1\tau_2\tau_4$ and $\tau_2\tau_3\tau_4$ (Fig.\ref{f3_cont}), 
\noindent
In all other kinematic regions  $f_{\omega_3}^{phase}$ vanishes.
The common factor $N_{\omega_3}$ is found to be 
\begin{eqnarray}
N_{\omega_3} = 2\sin(\pi\omega_{24})\sin(\pi\omega_{21})  \left( \frac{|s_1|}{\mu^2}\right)^{\omega_1}
\left( \frac{|s_2|}{\mu^2}\right)^{\omega_2}\left(\frac{|s_3|}{\mu^2}\right)^{\omega_3}\left(\frac{|s_4|}{\mu^2}\right)^{\omega_4}.
\end{eqnarray}
A comment is in place about the second line in  (4.22) which is proportional to $c_1-c_2$. As we will show in a few moments, the combination $c_1-c_2$ belongs to the "long" cut in the $\omega_2$ and $\omega_3$ channel. The fact that this combination also participates in our calculations of the "short" cut hints at the fact that the "long" cut contribution may contain terms which have the $\omega$ plane singularity structure of the "short" cuts, i.e. there is a mixing between the different Regge cuts. We will come back to this question in a forthcoming paper.   

An analogous discussion applies to the "short" cut in the $f_{\omega_2}$ channel (Fig.\ref{f2_cont}),
\begin{eqnarray}
N_{\omega_2} f_{\omega_2}^{phase}\;&=&\;\sin(\pi \omega_{3c})\left\{\frac{1}{2} d_1 + \frac{1}{2} d_2 - d_3\right\} + \sin(\pi \omega_{4c})\left\{b_1 - b_2\right\}-\\\nonumber &-& \frac{1}{\sin(\pi \omega_{41})}\left(\frac{1}{2}\sin(\pi \omega_{41})\sin(\pi \omega_{3c}) + \sin(\pi \omega_{13})\sin(\pi \omega_{4c}) \right)\left\{d_1 - d_2\right\}.
\end{eqnarray}
We continue the function $f_{\omega_2}^{phase}$ to those four different kinematic regions where it is nonzero:
\begin{figure}[H]
\centering
\epsfig{file=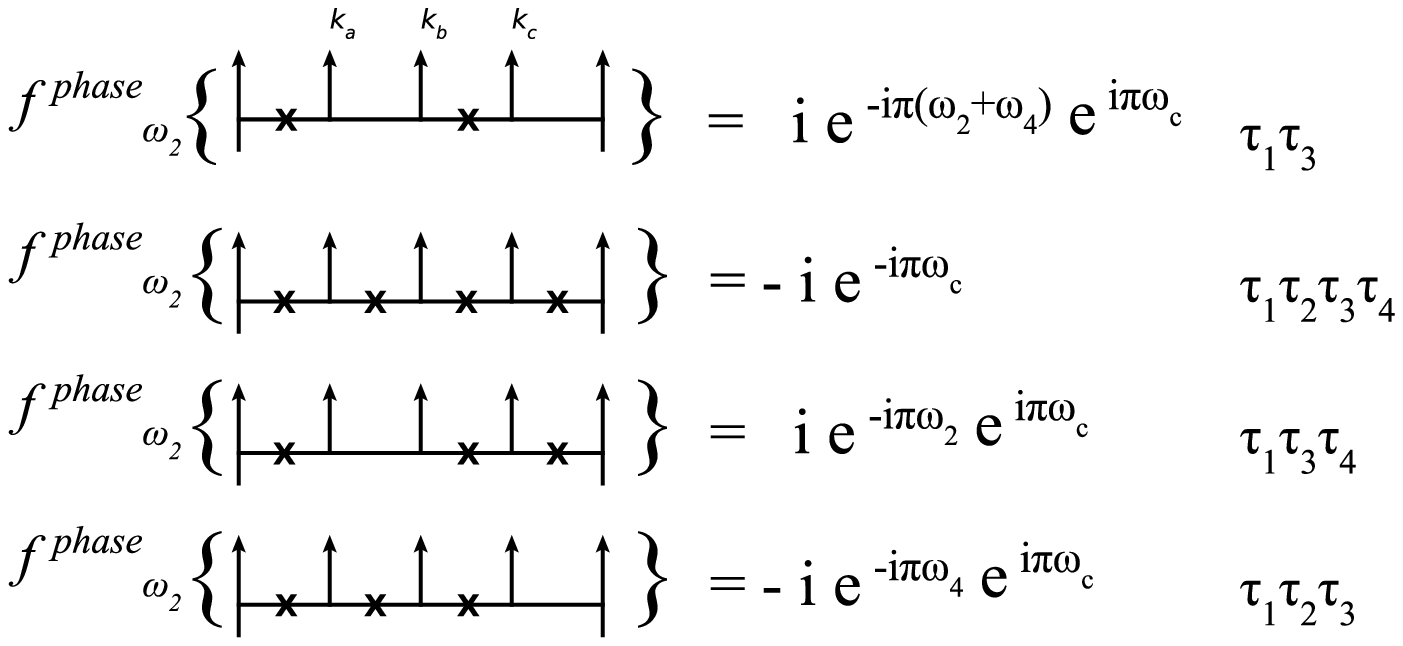}
\caption{Analytical continuation of $f_{\omega_2}$.}
\label{f2_cont}
\end{figure}
\noindent
with the common factor
\begin{eqnarray}
N_{\omega_2}= 2\sin(\pi\omega_{31})\sin(\pi\omega_{34})  \left( \frac{|s_1|}{\mu^2}\right)^{\omega_1}
\left( \frac{|s_2|}{\mu^2}\right)^{\omega_2}\left(\frac{|s_3|}{\mu^2}\right)^{\omega_3}\left(\frac{|s_4|}{\mu^2}\right)^{\omega_4}.
\end{eqnarray} 
In all other kinematic regions we have $f^{phase}_{\omega_2}\;=\;0$.

Next we turn to the "long" Regge cut in the $\omega_2$ and $\omega_3$ channels simultaneously. This cut is contained in the first two terms of the triplets $c_1$, $c_2$, $d_1$, $d_2$ of Fig.\ref{f3_to_w3all3} with the corresponding phases $c_1$, $c_2$, $d_1$, and $d_2$ in Eq.\eqref{ccuts} and Eq.\eqref{dcuts}. Repeating our line of arguments, we first consider the region where all energies are positive: since all $c_i$ are proportional to $e^{-i\pi \omega_3}$, all $d_i$ proportional to  $e^{-i\pi \omega_2}$, the coefficient of $c_1$ has to be opposite equal to that  $c_2$, and similarly for $d_1$ and $d_2$. Turning to the region $\tau_1\tau_2\tau_3\tau_4$,  the phases of $c_1-c_2$ are 
\be
c_1-c_2= 2i e^{-i\pi \omega_2} \sin (\pi \omega_{14}).
\ee
We take the following linear combination,
\be
\sin (\pi \omega_{3x})\left\{c_1 - c_2\right\} + \sin(\pi \omega_{2x})\left\{d_1 - d_2\right\}
= 2i  e^{-i\pi \omega_x} \sin (\pi \omega_{14}) \sin (\pi \omega_{32})
\ee
with $x=a,b,c$. Obviously, $x_b$ would be a symmetric choice; however the singular term in the Regge pole contribution (Appendix A) has no phase $e^{-i\pi \omega_b}$, and therefore this ansatz for the Regge cut cannot be used to subtract for the subtraction. Instead, we take the linear combination of two contributions, 
\begin{eqnarray}
N_{\omega_2 \omega_3}  f^{a;phase}_{(\omega_2 \omega_{3})}\;=\;\sin (\pi \omega_{3a})\left\{c_1 - c_2\right\} + \sin(\pi \omega_{2a})\left\{d_1 - d_2\right\}
\end{eqnarray}
and
\begin{eqnarray} 
N_{\omega_2 \omega_3}  f^{c;phase}_{(\omega_2 \omega_{3})}\;=\;\sin (\pi \omega_{3c})\left\{c_1 - c_2\right\} + \sin(\pi \omega_{2c})\left\{d_1 - d_2\right\},
\end{eqnarray}
and in the combination $A f^{a;phase}_{(\omega_2 \omega_{3})} + C f^{c;phase}_{(\omega_2 \omega_{3})}$ we will 
determine real valued coefficients $A=\delta f^{a}_{\omega_2 \omega_3}$ and $C=\delta f^{c}_{\omega_2 \omega_3}$ which subtract the singular part of the Regge pole contribution.

\begin{figure}[H]
\centering
\epsfig{file=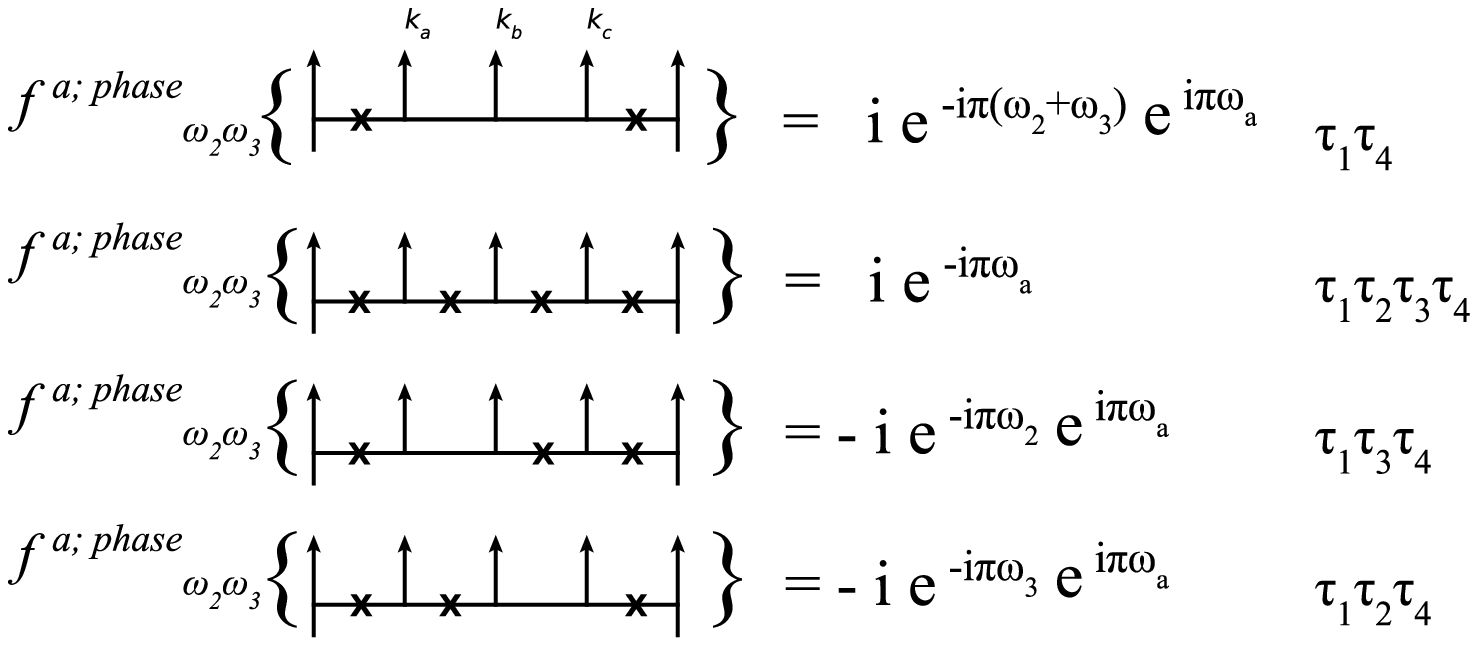}
\caption{Analytical continuation of $f^{a:phase}_{\omega_2\omega_3}$.}
\label{fa_cont}
\end{figure}

\begin{figure}[H]
\centering
\epsfig{file=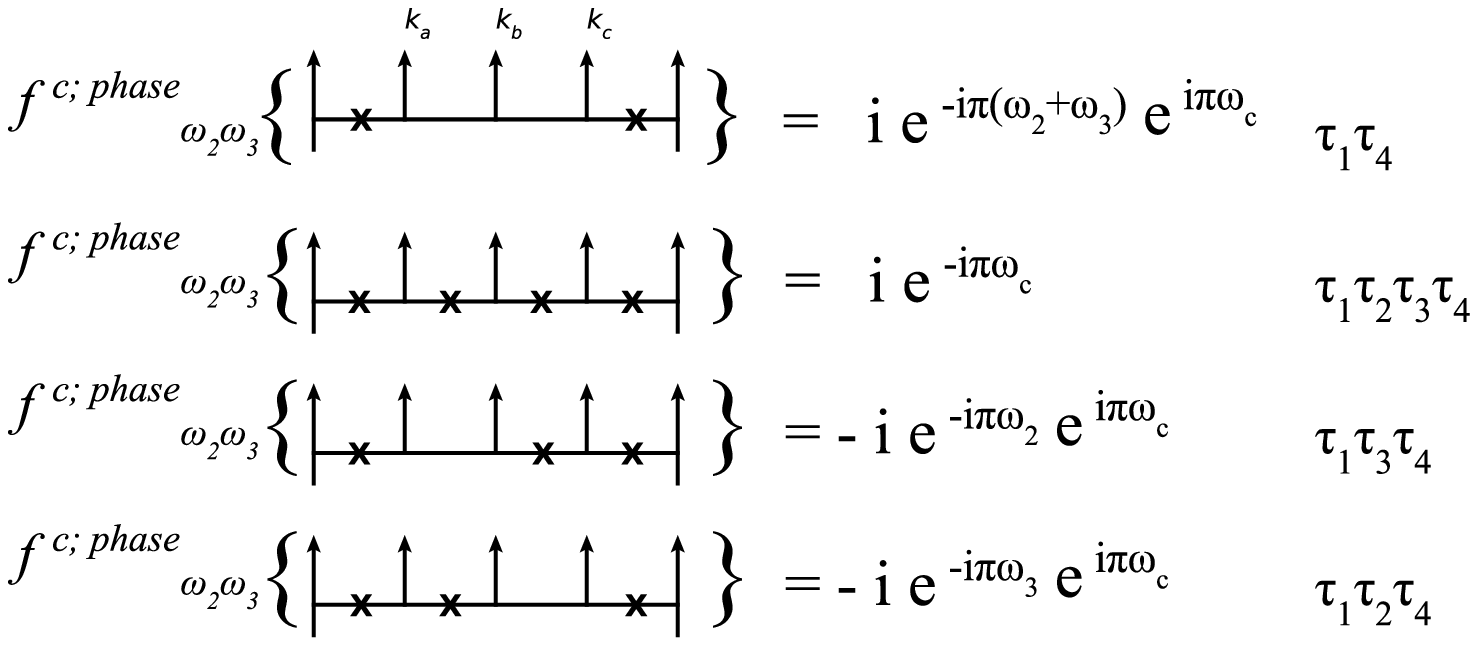}
\caption{Analytical continuation of $f^{c;phase}_{\omega_2\omega_3}$.}
\label{fc_cont}
\end{figure}

Let us first study the other kinematic regions.  The functions $f^{a;phase}_{\omega_2 \omega_{3}}$ and $f^{c;phase}_{\omega_2 \omega_{3}}$ have nonzero values in four particular kinematic regions (Fig.\ref{fa_cont}, 
(Fig.\ref{fc_cont})
\noindent
The common factor is the same for $f^{a;phase}_{\omega_2\omega_3}$ and for $f^{c;phase}_{\omega_2\omega_3}$:
\begin{eqnarray}
N_{\omega_2 \omega_3} = 2\sin(\pi\omega_{14})\sin(\pi\omega_{32})  \left( \frac{|s_1|}{\mu^2}\right)^{\omega_1}
\left( \frac{|s_2|}{\mu^2}\right)^{\omega_2}\left(\frac{|s_3|}{\mu^2}\right)^{\omega_3}\left(\frac{|s_4|}{\mu^2}\right)^{\omega_4}.
\end{eqnarray} 
For all other possible configuration of analytical continuation, the result is zero. Thus, the "long" cut contributes only to these four particular kinematic regions. We combine these two terms 
\begin{eqnarray}
\Delta f_{\omega_2\omega_3}\;=\; \delta f^{a}_{\omega_2 \omega_3} f^{a;phase}_{\omega_2\omega_3}
+ \delta f^{c}_{\omega_2 \omega_3} f^{c;phase}_{\omega_2\omega_3}
\end{eqnarray}
with real coefficients $\delta f^{a}_{\omega_2 \omega_3}$ and $\delta f^{c}_{\omega_2 \omega_3}$,  and we find for the different regions \footnote{We omitted the subscript `$\omega_2\omega_3$' of the $\delta f^{a,c}$ in the figure for the sake of simplicity.} (Fig.\ref{fd_cont}):
\begin{figure}[H]
\centering
\epsfig{file=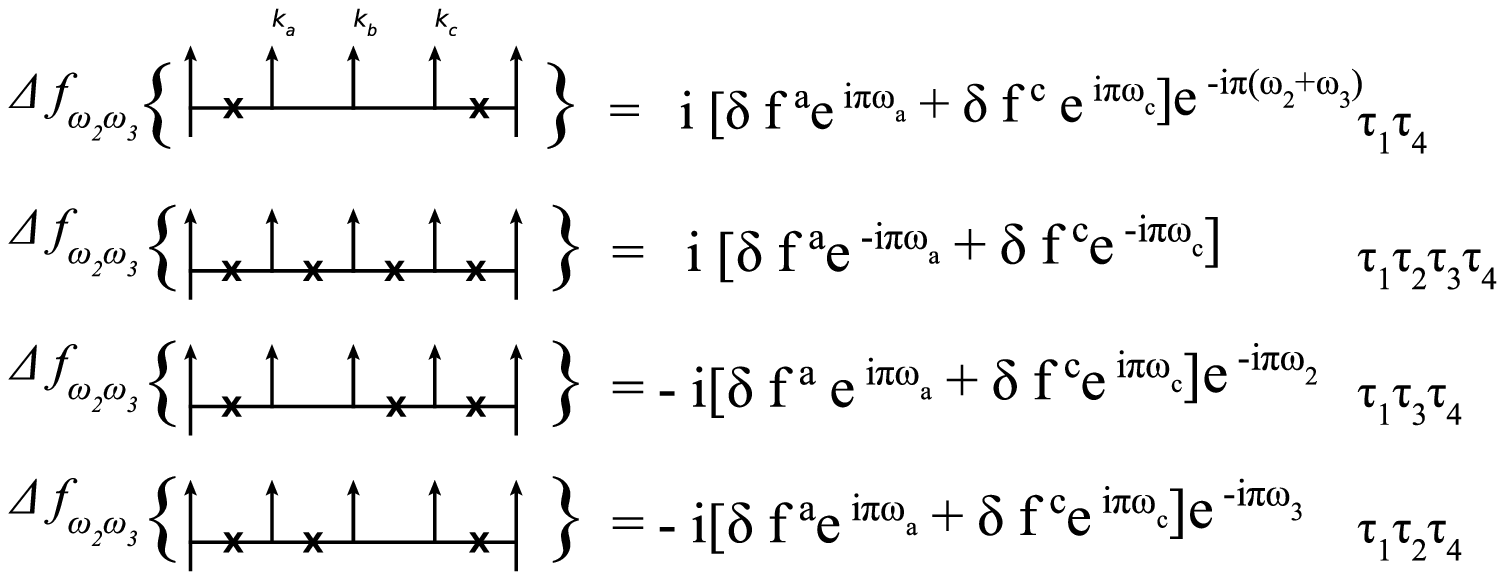}
\caption{Analytical continuation of $\Delta f_{\omega_2\omega_3}$.}
\label{fd_cont}
\end{figure}
\noindent
It is  remarkable that the square bracket is the same in all four cases, up to complex conjugation of the phases. Below we will determine the coefficients $\delta f^{a}_{\omega_2 \omega_3}$ and $\delta f^{c}_{\omega_2 \omega_3}$. Summarizing this subsection, we have determined coefficients of the partial waves $a_1,...d_3$ which, for all those  kinematic region which  contains Regge cuts, can be combined to give a "good" phase structure.  Returning to  Eq.\eqref{f_omega}, we have determined the normalization factors $N$ and the phases $f_{\omega}^{phase}$. In the following we still have to calculate the coefficients $\delta f_{\omega}$, and we have to show that our ansatz matches the phases of the singular pieces of the Regge pole terms (studied in Sec. II) and thus allows us to absorb these singularities by the Regge cuts. 

%%%%%%%%%%%%%%%%%%%%%%%%%%%%%%%%%%%%%%%%%%%%%%%%%%%%%%%%%%%%%%%%%%%%%%%%%%%%%%%%%%%%%%%%%%%%%%%%%%%%%%%%%
\subsection{Redefinitions of Regge pole terms: subtractions}

Let us now turn to the subtraction procedure. Figs.\ref{f3_cont}, \ref{f2_cont}, \ref{fd_cont} show the kinematic regions in which the different Regge cuts, $f_{\omega_3}$, $f_{\omega_2}$, and $f_{\omega_2 \omega_3 }$, contribute. There are two regions ($\tau_2\tau_4$ and $\tau_2\tau_3\tau_4$) in which only $f_{\omega_3}$ contributes, two regions ($\tau_1\tau_3$ and $\tau_1\tau_2\tau_3$) where only $f_{\omega_2}$ is nonzero, and one region ($\tau_1\tau_4$) where only the "long" cut appears. In the remaining three regions we have combinations of several Regge cuts. In particular, the region $\tau_1\tau_2\tau_3\tau_4$  sees all cut contributions. We begin with the "short" cut  $f_{\omega_3}$: from the region $\tau_2\tau_4$ we determine the subtraction  $\delta f_{\omega_3}$ which then fixes the subtractions in all regions listed in Fig.\ref{f3_cont}. Similarly,  $\delta f_{\omega_2}$ is obtained from the region $\tau_1\tau_3$ and will be used  in all regions listed in Fig.\ref{f2_cont}. Finally, in the region  $\tau_1\tau_2\tau_3\tau_4$ we can fix the remaining subtraction, $\delta f_{\omega_2 \omega_3 }$. 

We begin with the region  $\tau_2\tau_4$ where only the "short" cut in the $t_3$ channel contributes. From Appendix A we read off the Regge pole contribution in the region  $\tau_2\tau_4$,
\be
f_{pole}^{\tau_2 \tau_4} = e^{-i\pi\left(\omega_1+\omega_3\right)}e^{i\pi\omega_a}
\left(\cos (\pi \omega_{bc})+  \left[ e^{i\pi\left( \omega_b+\omega_c\right)} -\cos(\pi \omega_{bc}) - 2ie^{i\pi\omega_3} \frac{\sin(\pi\omega_b) \sin(\pi\omega_c) }{\sin (\pi\omega_3)} \right] \right).
\label{pole-tau2-tau4}
\ee
The square bracket expression on the rhs can also be written as [Eq.\eqref{2to4twist13a} and Eq.\eqref{2to4twist13}]
\be
[...]  =  +i\sin (\pi (\omega_b + \omega_c)) -2i \frac{\cos(\pi\omega_3)\sin (\pi \omega_b) \sin (\pi \omega_c)}{\sin (\pi \omega_3)},
\label{f_2-subtraction}
\ee 
which shows that it is purely imaginary and can also be written as 
\be
[...]=
-  \left[ e^{- i\pi\left( \omega_b+\omega_c\right)} -\cos(\pi \omega_{bc}) + 2ie^{-i\pi\omega_3} \frac{\sin(\pi\omega_b) \sin(\pi\omega_c) }{\sin (\pi\omega_3)} \right].
\ee
Thus the phase structure of the second part of Eq.\eqref{pole-tau2-tau4} is the same as that of the cut contribution $f^{phase}_{\omega_3}$ in the first line of Fig.\ref{f3_cont}, and we define the subtraction term as follows:
\ba
\delta  f_{\omega_3}&=&i \left[e^{i\pi\left( \omega_b+\omega_c\right)} -\cos (\pi \omega_{bc}) - 2ie^{i\pi\omega_3} \frac{\sin(\pi\omega_b) \sin (\pi\omega_c) }{\sin(\pi\omega_3)}\right]\nonumber\\ 
&=&-i \left[e^{-i\pi\left( \omega_b+\omega_c\right)} -\cos (\pi \omega_{bc}) + 2ie^{-i\pi\omega_3} \frac{\sin(\pi\omega_b) \sin (\pi\omega_c) }{\sin(\pi\omega_3)}\right]\nonumber\\
&=&- \left[ \sin (\pi (\omega_b + \omega_c)) -2 \frac{\cos(\pi\omega_3)\sin (\pi \omega_b) \sin (\pi \omega_c)}{\sin (\pi \omega_3)}\right].
\label{3-subtraction}
\ea
Having fixed the subtraction $\delta f_{\omega_3}$ in the $\tau_2\tau_4$ region, we know the subtraction for all kinematic regions in which the $\omega_3$-cut appears (these regions are listed in Fig.\ref{f3_cont}).  In our generating function we therefore have the following contributions\footnote{Note that here we follow our convention that terms promotional to an odd number of factors $\tau$ receive an additional minus sign.} 
\ba
-\left[ \tau_2 \tau_4 e^{-i\pi\left(\omega_1+\omega_3\right)}e^{i\pi\omega_a} - \tau_1\tau_2 \tau_4  e^{-i\pi \omega_3}e^{i\pi\omega_a} \right]\left[e^{i\pi\left( \omega_b+\omega_c\right)} -\cos (\pi \omega_{bc}) - 2ie^{i\pi\omega_3} \frac{\sin(\pi\omega_b) \sin (\pi\omega_c) }{\sin(\pi\omega_3)}\right]\nonumber\\
+ \left[ -\tau_2 \tau_3 \tau_4 e^{-i\pi \omega_1}e^{i\pi\omega_a} + \tau_1\tau_2 \tau_3 \tau_4  e^{-i\pi\omega_a} \right] \left[e^{-i\pi\left( \omega_b+\omega_c\right)} -\cos (\pi \omega_{bc}) + 2ie^{-i\pi\omega_3} \frac{\sin(\pi\omega_b) \sin (\pi\omega_c) }{\sin(\pi\omega_3)}\right].\nonumber\\
\label{omega3-cut-subtraction}
\ea
For the regions $\tau_2 \tau_4$ and  $\tau_2\tau_3 \tau_4$ these are the only subtractions, and by subtracting the corresponding parts of Eq.\eqref{omega3-cut-subtraction} from their Regge pole terms (Appendix A), all unwanted singular terms must cancel. Indeed, for the region $\tau_2 \tau_4$ we find
\ba
f_{ren; pole}^{\tau_2 \tau_4}  &=& f_{pole}^{\tau_2 \tau_4} + i e^{-i\pi \left( \omega_1+\omega_3 \right) } e^{i\pi\omega_a}  \delta f_{\omega_3} \nonumber \\
&=& e^{-i\pi\left( \omega_1+\omega_3\right)} e^{i\pi\omega_a} \cos (\pi \omega_{bc}),
\ea 
which consists of a phase factor and a conformal invariant expression: the latter will be called the "conformal Regge pole".  
Similarly, in the region $\tau_2\tau_3\tau_4$, together with the Regge pole contribution from Appendix A which we write as
\be
f_{pole}^{\tau_2 \tau_3 \tau_4}= -e^{-i\pi\omega_1}e^{i\pi\omega_a } \left( \cos (\pi \omega_{bc}) + \left[ e^{-i\pi (\omega_b+\omega_c)} -\cos (\pi \omega_{bc}) + 2ie^{-i\pi\omega_3}\frac{\sin(\pi\omega_b)\sin(\pi\omega_c)}{\sin(\pi\omega_3)} \right] \right)
\ee
we obtain 
\ba
f_{ren; pole}^{\tau_2 \tau_3 \tau_4}  &=& f_{pole}^{\tau_2 \tau_3 \tau_4} - i e^{-i\pi  \omega_1  } e^{i\pi\omega_a} \delta f_{\omega_3} \nonumber \\
&=& - e^{-i\pi \omega_1} e^{i\pi\omega_a} \cos (\pi \omega_{bc}).
\ea 
This defines our renormalized pole contribution in the region $\tau_2\tau_3\tau_4$.
The other two regions, $\tau_1 \tau_2 \tau_4$ and $\tau_1 \tau_2 \tau_3 \tau_4$, receive contributions also from  the "long" cut. They will be discussed further below.

A similar discussion applies to the symmetric region  $\tau_1\tau_3$  which is used 
to calculate the subtraction contained in $f_{\omega_2}$:
\ba
\delta f_{\omega_2}
&=&i \left[e^{i\pi\left( \omega_a+\omega_b\right)} -\cos (\pi \omega_{ab}) - 2ie^{i\pi\omega_2} \frac{\sin(\pi\omega_a) \sin (\pi\omega_b) }{\sin (\pi\omega_2)}\right] \nonumber \\
&=&- i \left[e^{-i\pi\left( \omega_a+\omega_b\right)} -\cos (\pi \omega_{ab}) + 2ie^{-i\pi\omega_2} \frac{\sin(\pi\omega_a) \sin (\pi\omega_b) }{\sin (\pi\omega_2)}\right] \nonumber \\
&=&- \left[ \sin (\pi (\omega_a + \omega_b)) -2 \frac{\cos(\pi\omega_2)\sin (\pi \omega_a) \sin (\pi \omega_b)}{\sin (\pi \omega_2)}\right].\nonumber\\
\label{2-subtraction}
\ea
From Fig.\ref{f2_cont} it follows that the same subtraction contributes also to the regions $\tau_1 \tau_2 \tau_3$, $\tau_1 \tau_3 \tau_4$, and  $\tau_1 \tau_2 \tau_3 \tau_4$. The analogue of Eq.\eqref{omega3-cut-subtraction} reads,
\ba
-\left[ \tau_1 \tau_3 e^{-i\pi\left(\omega_2+\omega_4\right)}e^{i\pi\omega_c} - \tau_1\tau_3 \tau_4  e^{-i\pi \omega_2}e^{i\pi\omega_c} \right]\left[e^{i\pi\left( \omega_a+\omega_b\right)} -\cos (\pi \omega_{ab}) - 2ie^{i\pi\omega_2} \frac{\sin(\pi\omega_a) \sin (\pi\omega_b) }{\sin(\pi\omega_2)}\right]\nonumber\\
+ \left[ -\tau_1 \tau_2 \tau_3 e^{-i\pi \omega_1}e^{i\pi\omega_c} + \tau_1\tau_2 \tau_3 \tau_4  e^{-i\pi\omega_c} \right] \left[e^{-i\pi\left( \omega_a+\omega_b\right)} -\cos (\pi \omega_{ab}) + 2ie^{-i\pi\omega_2} \frac{\sin(\pi\omega_a) \sin (\pi\omega_b) }{\sin(\pi\omega_2)}\right],\nonumber\\
\label{omega2-cut-subtraction}
\ea
and the renormalized Regge poles in the  regions  $\tau_1\tau_3$ and $\tau_1 \tau_2 \tau_3$ have the form:
\ba
f_{ren; pole}^{\tau_1 \tau_3}  &=& f_{pole}^{\tau_1 \tau_3} + i e^{-i\pi \left( \omega_2+\omega_4 \right) } e^{i\pi\omega_c}  \delta f_{\omega_2} \nonumber \\
&=& e^{-i\pi\left( \omega_2+\omega_4\right)} e^{i\pi\omega_c} \cos (\pi \omega_{ab}),
\ea 
and
\ba
f_{ren; pole}^{\tau_1 \tau_2 \tau_3}  &=& f_{pole}^{\tau_1 \tau_2 \tau_3} - i e^{-i\pi  \omega_4  } e^{i\pi\omega_c} \delta f_{\omega_2} \nonumber \\
&=& - e^{-i\pi \omega_4} e^{i\pi\omega_c} \cos (\pi \omega_{ab}).
\ea 
Finally we turn to the contributions of the "long" cut which contributes to the regions listed in Fig.\ref{fd_cont}. We start with the region $\tau_1 \tau_2\tau_3  \tau_4$: in this region all three cuts contribute. The subtractions contained in the two "short" cuts have already been determined before, and we can use these results for fixing the subtraction due to the "long" cut. We again begin with the Regge pole expression (from the Appendix A):  
\be 
f_{pole}^{\tau_1  \tau_2\tau_3\tau_4}=
e^{i\pi\left(-\omega_a+\omega_b-\omega_c\right)} - 2i\frac{\sin(\pi\omega_{2a})\sin(\pi\omega_b)\sin(\pi\omega_{3c})}{\sin(\pi\omega_2)\sin(\pi\omega_3)}.
\label{pole1234}
\ee
The subtractions from the "short" cuts, $\delta f_{\omega_3}$ and $\delta f_{\omega_2}$, have been defined above: $\delta f_{\omega_3}$ in Eq.\eqref{3-subtraction} and Eq.\eqref{omega3-cut-subtraction}, and $\delta f_{\omega_2}$ in Eq.\eqref{2-subtraction} and Eq.\eqref{omega2-cut-subtraction}. Before the subtraction due to the "long" cut, we have, 
\be
f_{pole}^{\tau_1  \tau_2\tau_3\tau_4} - i e^{-i \pi \omega_c} \delta f_{\omega_2 }-   i e^{-i \pi \omega_a}  \delta f_{\omega_3} 
%+ i e^{-i \pi \omega_a}  \delta f^{a}_{\omega_2 \omega_3} + i e^{-i \pi \omega_c}  \delta f^{c}_{\omega_2 \omega_3}, 
\label{1234-renorm}
\ee
which contains a double pole term  $\sim 1/\left( \sin (\pi \omega_2) \sin (\pi \omega_3) \right)$ [from $f_{pole}^{\tau_1  \tau_2\tau_3\tau_4}$ in Eq.\eqref{pole1234}] and single poles $\sim 1/\sin(\pi \omega_i)$ (i=1,2) (from  $\delta f_{\omega_3}$ and $\delta f_{\omega_2}$). We now use the freedom of having another subtraction connected with the "long" cut, $f^{a,c}_{\omega_2 \omega_3}$. We chose these remaining subtractions  $\delta f^{a,c}_{\omega_2 \omega_3}$  in such a way that they remove all double poles  $\sim 1/\left( \sin (\pi \omega_2) \sin (\pi \omega_3) \right) $, all single poles $\sim 1/\sin(\pi \omega_i)$ (i=1,2), and make the resulting expression conformally invariant. This leads to,
\ba
\Delta  f_{\omega_2 \omega_3}&=& \left\{
%e^{i\pi\left(-\omega_a+\omega_b-\omega_c\right)} 
- 2\frac{\sin(\pi\omega_{2a})\sin(\pi\omega_b)\sin(\pi\omega_{3c})}{\sin(\pi\omega_2)\sin(\pi\omega_3)} - \right.\nonumber\\
&&\;\;- e^{-i\pi\omega_a}i \left[e^{-i\pi\left(\omega_b+\omega_c\right)} - \cos(\pi\omega_{bc})+ 2ie^{-i\pi\omega_3}\frac{\sin(\pi\omega_b)\sin(\pi\omega_c)}{\sin(\pi\omega_3)}\right]-\nonumber\\
&&\;\ \left.- e^{-i\pi\omega_c}i\left[e^{-i\pi\left(\omega_a+\omega_b\right)} - \cos(\pi\omega_{ab})+ 2ie^{-i\pi\omega_2}\frac{\sin(\pi\omega_a)\sin(\pi\omega_b)}{\sin(\pi\omega_2)}\right]
\right\}. 
\label{f_{23}-subtraction}
\ea
Here we remind that, according to Eq.\eqref{f_2-subtraction}, the square brackets in the second and third rows are purely imaginary. The first term can also be written in the form 
\ba
- 2\frac{\sin(\pi\omega_{2a})\sin(\pi\omega_b)\sin(\pi\omega_{3c})}{\sin(\pi\omega_2)\sin(\pi\omega_3)}= \hspace{3cm}\\
\left(e^{-i\pi \omega_a} \frac{\sin (\pi \omega_c)}{\sin(\pi \omega_{ac})} +  e^{-i\pi \omega_c} \frac{\sin (\pi \omega_a)}{\sin(\pi \omega_{ca})}\right) 2 \frac{\sin(\pi\omega_{2a})\sin(\pi\omega_b)\sin(\pi\omega_{3c})}{\sin(\pi\omega_2)\sin(\pi\omega_3)}.\nonumber
\ea
Inserting this into Eq.\eqref{f_{23}-subtraction} one sees that, in fact, $ \Delta f_{\omega_2 \omega_3}$ can be written as:  
\be
\Delta f_{\omega_2 \omega_3}= \delta f^{a}_{\omega_2 \omega_3} e^{-i\pi \omega_a} 
+ \delta f^{c}_{\omega_2 \omega_3} e^{-i\pi \omega_c}
\ee
with real coefficients $\delta f^{a}_{\omega_2 \omega_3}$ and $\delta f^{c}_{\omega_2 \omega_3}$. 

Having fixed the subtractions  due to the "long" cut,  $\delta f^{a}_{\omega_2 \omega_3}$ and $\delta f^{c}_{\omega_2 \omega_3}$, we must show that in all four kinematic regions in which the "long" cut is nonzero (Fig.\ref{fd_cont}), the unphysical singularities of the Regge pole contributions cancel. We  collect these subtractions by writing them as part of the generating function, 
\begin{eqnarray}\label{subtraction_step2}
&&\left\{e^{i\pi\left(\omega_a+\omega_b+\omega_c\right)}e^{-i\pi\left(\omega_2+\omega_3\right)} - 2i\frac{\sin(\pi\omega_a)\sin(\pi\omega_b)\sin(\pi\omega_c)}{\sin(\pi\omega_2)\sin(\pi\omega_3)}\right\}\tau_1\tau_4\\
&&\left\{e^{i\pi\left(-\omega_a+\omega_b-\omega_c\right)} - 2i\frac{\sin(\pi\omega_{2a})\sin(\pi\omega_b)\sin(\pi\omega_{3c})}{\sin(\pi\omega_2)\sin(\pi\omega_3)}- \right.\nonumber\\
&&\;\;-e^{-i\pi\omega_a}\left[e^{-i\pi\left(\omega_b+\omega_c\right)} - \cos(\omega_{bc})+ 2ie^{-i\pi\omega_3}\frac{\sin(\pi\omega_b)\sin(\pi\omega_c)}{\sin(\pi\omega_3)}\right]\nonumber\\
&&\;\;\left.-e^{-i\pi\omega_c}\left[e^{-i\pi\left(\omega_a+\omega_b\right)} - \cos(\omega_{ab})+ 2ie^{-i\pi\omega_2}\frac{\sin(\pi\omega_a)\sin(\pi\omega_b)}{\sin(\pi\omega_2)}\right]
\right\}\tau_1\tau_2\tau_3\tau_4\nonumber\\
&-&\left\{e^{i\pi\left(-\omega_a+\omega_b+\omega_c\right)}e^{-i\pi\omega_3} - 2i\frac{\sin(\pi\omega_{2a})\sin(\pi\omega_b)\sin(\pi\omega_c)}{\sin(\pi\omega_2)\sin(\pi\omega_3)} - \right.\nonumber\\
&&\;\;-\left.e^{-i\pi\omega_3}e^{i\pi\omega_a}\left[e^{i\pi\left(\omega_b+\omega_c\right)} - \cos(\omega_{bc}) - 2ie^{i\pi\omega_3}\frac{\sin(\pi\omega_b)\sin(\pi\omega_c)}{\sin(\pi\omega_3)}\right]\right\}\tau_1\tau_2\tau_4\nonumber\\
&-&\left\{e^{i\pi\left(\omega_a+\omega_b-\omega_c\right)}e^{-i\pi\omega_2} - 2i\frac{\sin(\pi\omega_a)\sin(\pi\omega_b)\sin(\pi\omega_{3c})}{\sin(\pi\omega_2)\sin(\pi\omega_3)} - \right.\nonumber\\
&&\;\;-\left.e^{-i\pi\omega_2}e^{i\pi\omega_c}\left[e^{i\pi\left(\omega_a+\omega_b\right)} - \cos(\omega_{ab}) - 2ie^{i\pi\omega_2}\frac{\sin(\pi\omega_a)\sin(\pi\omega_b)}{\sin(\pi\omega_2)}\right]\right\}\tau_1\tau_3\tau_4.\nonumber\\
\end{eqnarray} 
It is now a matter of straightforward algebra to calculate the conformal Regge poles for the four different kinematic regions. For the region $\tau_1\tau_2\tau_3\tau_4$ we return to Eq.\eqref{1234-renorm} and find: 
\ba
f_{ren;pole}^{\tau_1  \tau_2\tau_3\tau_4}&=& f_{pole}^{\tau_1  \tau_2\tau_3\tau_4} - i e^{-i \pi \omega_c} \delta f_{\omega_2 }- i e^{-i \pi \omega_a}  \delta f_{\omega_3} + i \Delta f_{\omega_2 \omega_3}\nonumber\\
&=& e^{i \pi(-\omega_a+\omega_b-\omega_c)}.
\ea
Here the "conformal Regge pole" equals unity. In the same way we find for the other regions,
\ba
f_{ren;pole}^{\tau_1  \tau_4}&=& f_{pole}^{\tau_1  \tau_4} + 
i e^{-i\pi (\omega_2+\omega_3)} \Delta f_{\omega_2  \omega_3}\nonumber\\
&=& e^{-i\pi (\omega_2+\omega_3)}  e^{i \pi \omega_b} \cos (\pi \omega_{ac}),
\ea
\ba
f_{ren;pole}^{\tau_1 \tau_2 \tau_4}&=& f_{pole}^{\tau_1  \tau_2 \tau_4}+  ie^{-i\pi \omega_{3}}\delta f_{\omega_3} -
i e^{-i\pi \omega_{3}} \Delta f_{\omega_2  \omega_3}\nonumber\\
&=& - e^{-i\pi \omega_{3}}  e^{i\pi \omega_{c}}  \cos (\pi \omega_{ab}),
\ea
and
\ba
f_{ren;pole}^{\tau_1  \tau_3 \tau_4}&=& f_{pole}^{\tau_1  \tau_3 \tau_4} - i e^{-i\pi \omega_{2}}\delta f_{\omega_2} -
i e^{-i\pi \omega_{2}} \Delta f_{\omega_2  \omega_3}\nonumber\\
&=&-  e^{-i\pi \omega_{2}} e^{i\pi \omega_{a}}  \cos (\pi \omega_{bc}).
\ea

%%%%%%%%%%%%%%%%%%%%%%%%%%%%%%%%%%%%%%%%%%%%%%%%%%%%%%%%%%%%%%%%%%%%%%%%%%%%%%%%%%%%%%%%%%%%%%%%%%%%%%%%%

\subsection{Predictions for the remainder function of the $2\rightarrow 5$ amplitude}
Let us summarize our results for those eight kinematic regions for which the Regge pole terms need to be renormalized. This are also the regions which contain Regge cuts. We again use our notation of a generating function and write for the scattering amplitude $A$: 
\be
A\;=\; A_0 + A_1 \tau_1 + ...+ A_{12} \tau_1 \tau_2 +... + A_{1234} \tau_1\tau_2\tau_3\tau_4.
\ee
Here each term proportional to $\tau_i...\tau_j$ is written as a product of the BDS prediction and a remainder function:  
\be
          A_{i..j} = A_{BDS; i..j} R_{\tau_i ...\tau_j} ,
\ee
and in Sec. III it has been shown that the BDS amplitude $A_{BDS;i..j}$ can be written as the product of a real part, a kinematic phase factor, and a second phase factor $e^{i\delta_{i...j}}$, where  the conformal invariant $\delta_{i...j}$ result from the $Li_2$ functions and represent the one-loop approximations to Regge cut contributions,
\be
A_{BDS;i..j} =\pm |A_{BDS;i..j}| e^{i\varphi_{i...j}} e^{i\delta_{i...j}}.
\ee
In the following Fig.\ref{table_mobius} we list the phase factors $e^{i\varphi_{i...j}}$,
\begin{figure}[H]
\centering
\epsfig{file=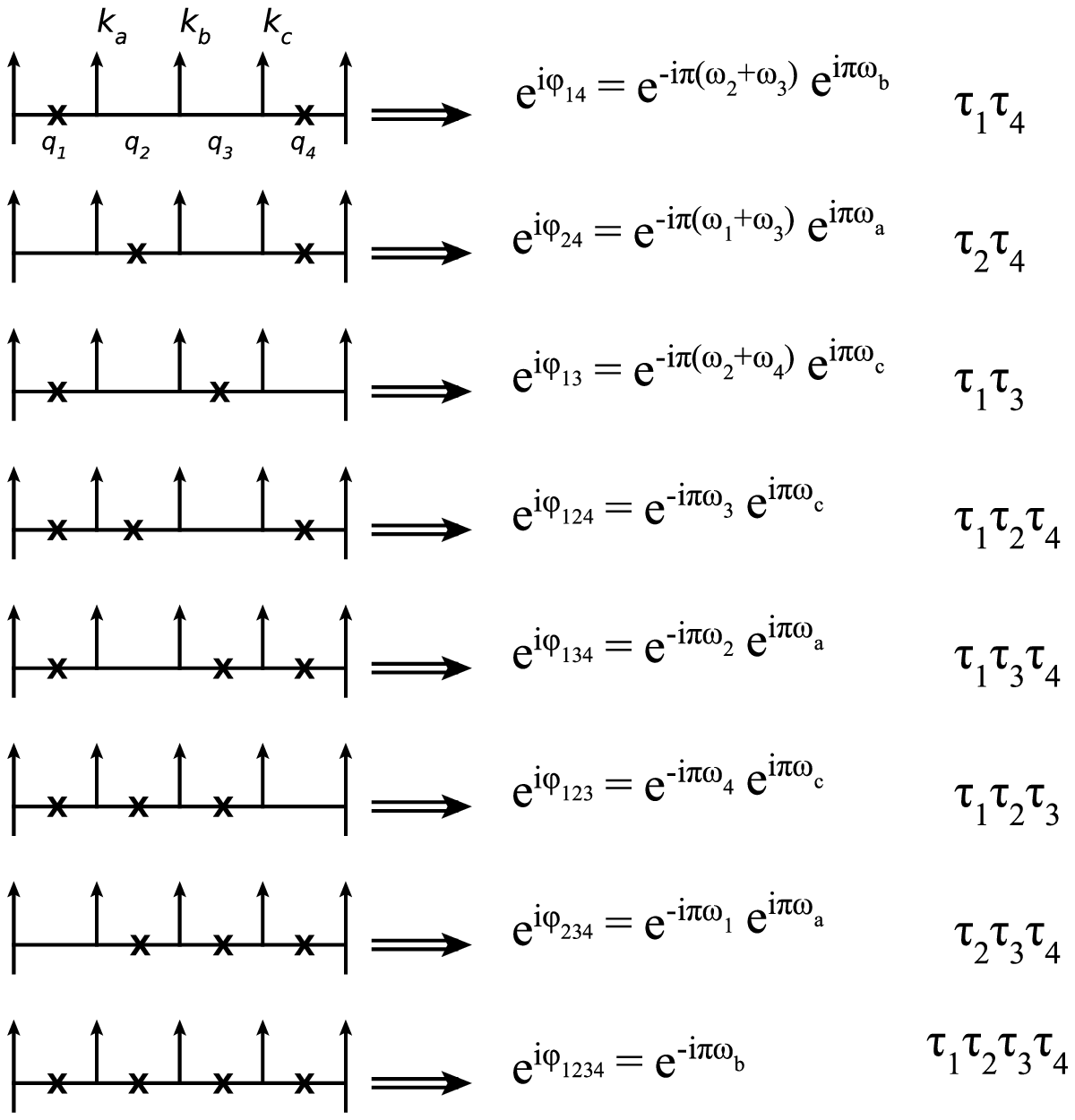,scale=1.25}
\caption{Phase factors $\varphi_{i...j}$ of the $2\rightarrow 5$ amplitude.}
\label{table_mobius}
\end{figure}
\noindent
Next, we collect the phases $\delta_{i...j}$:
\begin{eqnarray}
\delta_{14}\;&=&\;\pi\frac{\gamma_K}{4}\ln\left(\frac{|k_a||k_c||q_1||q_4|}{|k_a+k_b+k_c|^2|q_2||q_3|}\right)\\
\delta_{24}\;&=&\;\pi\frac{\gamma_K}{4}\ln\left(\frac{|k_b||k_c||q_2||q_4|}{|k_b+k_c|^2|q_3|^2}\right)\nonumber\\
\delta_{13}\;&=&\;\pi\frac{\gamma_K}{4}\ln\left(\frac{|k_a||k_b||q_1||q_3|}{|k_a+k_b|^2|q_2|^2}\right)\nonumber\\
\delta_{124}\;&=&\;\pi\frac{\gamma_K}{4}\ln\left(\frac{|k_a+k_b+k_c|^2|k_a||q_2|^2|q_3|}{|k_b+k_c|^2|k_b||q_1|^3}\right)\nonumber\\
\delta_{134}\;&=&\;\pi\frac{\gamma_K}{4}\ln\left(\frac{|k_a+k_b+k_c|^2|k_c||q_2||q_3|^2}{|k_a+k_b|^2|k_b||q_4|^3}\right)\nonumber\\
\delta_{123}\;&=&\;\pi\frac{\gamma_K}{4}\ln\left(\frac{|k_a||k_b||q_1||q_3|}{|k_a+k_b|^2|q_2|^2}\right)\nonumber\\
\delta_{234}\;&=&\;\pi\frac{\gamma_K}{4}\ln\left(\frac{|k_b||k_c||q_2||q_4|}{|k_b+k_c|^2|q_3|^2}\right)\nonumber\\
\delta_{1234}\;&=&\;\pi\frac{\gamma_K}{4}\ln\left(\frac{|k_a+k_b|^2|k_b+k_c|^2|q_1||q_4|}{|k_a+k_b+k_c|^2|k_a||k_c||q_2||q_3|}\right).\nonumber
\end{eqnarray}
Finally, we collect the conformal invariant Regge pole and cut terms which have been calculated in the previous subsection and represent the main results of Sec. IV.
They define our predictions for the remainder function $R$, more precisely for the products 
$R_{\tau_i...\tau_j} e^{i\delta_{ij}}$:  
\begin{eqnarray}
\tau_1\tau_4: &&\;\;\;\cos(\pi\omega_{ac})+i\,\left( e^{i\pi\omega_{ba}} f^{a}_{\omega_2 \omega_3}+ e^{i\pi\omega_{bc}} f^{c}_{\omega_2 \omega_3} \right)\\
\tau_2\tau_4: &&\;\;\;\cos(\pi\omega_{bc})+i\,f_{\omega_{3}}\nonumber\\
\tau_1\tau_3: &&\;\;\;\cos(\pi\omega_{ab})+i\,f_{\omega_2}
\nonumber\\
\tau_1\tau_2\tau_4: &&\;\;\; - \cos(\pi\omega_{ab})-i\, e^{i\pi\omega_{ac}} f_{\omega_3}-i\, e^{i\pi\omega_{ac}} f^{a}_{\omega_2 \omega_3} -i\,  f^{c}_{\omega_2 \omega_3}\nonumber\\
\tau_1\tau_3\tau_4: &&\;\;\;- \cos(\pi\omega_{bc})-i\, e^{i\pi\omega_{ca}} f_{\omega_2}-i\, f^{a}_{\omega_2 \omega_3} -i\,  e^{i\pi\omega_{ca}}f^{c}_{\omega_2 \omega_3}\,\nonumber\\
\tau_1\tau_2\tau_3:&&\;\;\;-\cos (\pi \omega_{ab}) - i\, f_{\omega_2} \nonumber\\
\tau_2\tau_3\tau_4:&&\;\;\;-\cos (\pi \omega_{bc}) - i\, f_{\omega_3} \nonumber\\
\tau_1\tau_2\tau_3\tau_4:&&\;\;\; e^{i\pi \omega_{ba}} e^{i\pi \omega_{bc}} -i\, e^{i\pi\omega_{ba}} f_{\omega_3} 
-i\, e^{i\pi\omega_{bc}} f_{\omega_2} +i\, e^{i\pi\omega_{ba}}f^{a}_{\omega_2 \omega_3}+i\, e^{i\pi\omega_{bc}}f^{c}_{\omega_2 \omega_3}
\nonumber
\end{eqnarray}
The conformal invariant Regge cut terms $f_{\omega_2}$,  $f_{\omega_3}$, $f^{a,c}_{\omega_2\omega_3}$ contain, in addition to the subtraction terms  $\delta f_{\omega_2}$,  $\delta f_{\omega_3}$, $\delta f^{a,c}_{\omega_2\omega_3}$, respectively, which we have discussed in subsection IVC, the terms with Regge cut singularities. In this paper, we have not addressed yet the general structure of these amplitudes. This will be the subject of a forthcoming paper.  
%%%%%%%%%%%%%%%%%%%%%%%%%%%%%%%%%%%%%%%%%%%%%%%%%%%%%%%%%%%%%%%%%%%%%%%%%%%%%%%%%%%%%%%%%%%%%%%%%%%%
%%%%%%%%%%%%%%%%%%%%%%%%%%%%%%%%%%%%%%%%%%%%%%%%%%%%%%%%%%%%%%%%%%%%%%%%%%%%%%%%%%%%%%%%%%%%%%%%%%%%
%%%%%%%%%%%%%%%%%%%%%%%%%%%%%%%%%%%%%%%%%%%%%%%%%%%%%%%%%%%%%%%%%%%%%%%%%%%%%%%%%%%%%%%%%%%%%%%%%%%%
%%%%%%%%%%%%%%%%%%%%%%%%%%%%%%%%%%%%%%%%%%%%%%%%%%%%%%%%%%%%%%%%%%%%%%%%%%%%%%%%%%%%%%%%%%%%%%%%%%%%

\section{Conclusions}

In this paper we have addressed different aspects of scattering amplitudes in the multi-Regge region. Starting from Regge pole models that factorize in the kinematic region of positive energies, we have seen that after analytic continuation to other kinematic regions, terms with unphysical poles appear which need to be compensated by other terms. Specializing to the planar approximation of the conformal $\mathcal{N}=4$ SYM theory, we have studied the cases $2\to4$ and $2\to5$, and we have shown that it is possible to compute, in agreement with the analytic structure dictated by the Steinmann relations,  coefficients of Regge cut contributions which match the singular Regge pole pieces and thus can be used to absorb the singularities. We have outlined a "renormalization scheme" that consistently removes the singularities and leads to conformal invariance of the pole contribution.
 
Since most of this has been motivated by the goal of determining the remainder function $R_{n}$ in 
$\mathcal{N}=4$ SYM theory, we have systematically studied the predictions of the BDS formula in multi-Regge kinematics for the different kinematic regions, and compared them with our results for Regge pole models and Regge cuts. This has led  us to the definition of a remainder function that contains, apart from the Regge cut contribution, a conformal invariant Regge pole term. In this paper, we have not addressed the detailed structure of the Regge cut terms; this will be the content of a separate paper.   

In a future study we will extend our study to the case $2\to6$, which is expected to contain a new form of the Regge cut consisting of three Reggeized gluons.

\section{Acknowledgments}

We thank B. Basso, S. Caron-Huot, L. Dixon, V.S. Fadin, A. Prygarin, A. Sabio Vera, V. Schomerus, A. Sever, M. Spradlin, M. Sprenger, and A. Volovich for helpful discussions. The work of A.K. is supported by the Minerva Fellowship Program of Max Planck Gesellschaft.

\appendix
\newpage
\section{Explicit results of the $\tau$ expansion for the $n=5,6,7$ point amplitudes}

In this part we summarize the explicit coefficients of the $\tau$ expansions, $P_{2 \to n}$, for the cases $2\to3$, $2\to4$, $2\to5$, and $2\to6$. We start with the simplest case of $n=5$ amplitude and list all terms:
\begin{table}[H]
\begin{center}
\begin{tabular}{l r l}
\vspace{0.3cm}
$P_{2\rightarrow3}\;=\;$ & $e^{i\pi\omega_a}e^{-i\pi\left(\omega_1+\omega_2\right)}$ & (free term)  \\ 
\vspace{0.3cm}
& $-e^{i\pi\omega_a}e^{-i\pi\omega_2}$ & $\tau_1$ \\ 
\vspace{0.3cm}
& $-e^{i\pi\omega_a}e^{-i\pi\omega_1}$ & $\tau_2$\\ 
& $e^{-i\pi\omega_a}$ & $\tau_1\tau_2$.\\ 
\end{tabular}
\end{center}
\caption{All terms of the production amplitude $P_{2\rightarrow3}$.}\label{table1}
\end{table}\label{appA}
\noindent
%\newpage
Next we summarize the $n=6$ amplitude:
\begin{table}[H]
\begin{center}
\begin{tabular}{l r l}
\vspace{0.3cm}
$P_{2\rightarrow4}\;=\;$ & $e^{i\pi\left(\omega_a+\omega_b\right)}e^{-i\pi\left(\omega_1+\omega_2+\omega_3\right)}$ & (free term)  \\ 
\vspace{0.3cm}
& $-e^{i\pi\left(\omega_a+\omega_b\right)}e^{-i\pi\left(\omega_2+\omega_3\right)}$ & $\tau_1$ \\ 
\vspace{0.3cm}
& $-e^{i\pi\left(\omega_a+\omega_b\right)}e^{-i\pi\left(\omega_1+\omega_3\right)}$ & $\tau_2$\\ 
\vspace{0.3cm}
& $-e^{i\pi\left(\omega_a+\omega_b\right)}e^{-i\pi\left(\omega_1+\omega_2\right)}$ & $\tau_3$\\
\vspace{0.3cm}
& $e^{i\pi\left(\omega_b-\omega_a\right)}e^{-i\pi\omega_3}$ & $\tau_1\tau_2$\\ 
\vspace{0.3cm}
& $e^{i\pi\left(\omega_a-\omega_b\right)}e^{-i\pi\omega_1}$ & $\tau_2\tau_3$\\
\vspace{0.3cm}
& $e^{-i\pi\omega_2}\left[e^{i\pi\left(\omega_a+\omega_b\right)} - 2ie^{i\pi\omega_2}\frac{\sin(\pi\omega_a)\sin(\pi\omega_b)}{\sin(\pi\omega_2)} \right]$ & $\tau_1\tau_3$\\
\vspace{0.3cm}
& $-\left[e^{-i\pi\left(\omega_a+\omega_b\right)} + 2ie^{-i\pi\omega_2}\frac{\sin(\pi\omega_a)\sin(\pi\omega_b)}{\sin(\pi\omega_2)} \right]$ & $\tau_1\tau_2\tau_3$.
\end{tabular} 
\end{center}
\caption{All terms of the production amplitude $P_{2\rightarrow4}$.}\label{table2}
\end{table}
\newpage
Finally, the  coefficients of $n=7$ amplitude:
\begin{table}[H]
\begin{center}
\begin{tabular}{l r l}
\vspace{0.3cm}
$P_{2\rightarrow5}\;=\;$ & $e^{i\pi\left(\omega_a+\omega_b+\omega_c\right)}e^{-i\pi\left(\omega_1+\omega_2+\omega_3+\omega_4\right)}$ & (free term)  \\ 
\vspace{0.3cm}
& $-e^{i\pi\left(\omega_a+\omega_b+\omega_c\right)}e^{-i\pi\left(\omega_2+\omega_3+\omega_4\right)}$ & $\tau_1$ \\ 
\vspace{0.3cm}
&$-e^{i\pi\left(\omega_a+\omega_b+\omega_c\right)}e^{-i\pi\left(\omega_1+\omega_3+\omega_4\right)}$ & $\tau_2$\\ 
\vspace{0.3cm}
&$-e^{i\pi\left(\omega_a+\omega_b+\omega_c\right)}e^{-i\pi\left(\omega_1+\omega_2+\omega_4\right)}$ & $\tau_3$\\
\vspace{0.3cm}
&$-e^{i\pi\left(\omega_a+\omega_b+\omega_c\right)}e^{-i\pi\left(\omega_1+\omega_2+\omega_3\right)}$ & $\tau_4$\\
\vspace{0.3cm}
&$e^{-i\pi\left(\omega_a-\omega_b-\omega_c\right)}e^{-i\pi\left(\omega_3+\omega_4\right)}$ & $\tau_1\tau_2$\\ 
\vspace{0.3cm}
&$e^{-i\pi\left(\omega_2+\omega_4\right)}e^{i\pi\omega_c}\left[e^{i\pi\left(\omega_a+\omega_b\right)} - 2ie^{i\pi\omega_2}\frac{\sin(\pi\omega_a)\sin(\pi\omega_b)}{\sin(\pi\omega_2)} \right]$ & $\tau_1\tau_3$\\
\vspace{0.3cm}
&$\left[e^{i\pi\left(\omega_a+\omega_b+\omega_c\right)}e^{-i\pi\left(\omega_2+\omega_3\right)} - 2i\frac{\sin(\pi\omega_a)\sin(\pi\omega_b)\sin(\pi\omega_c)}{\sin(\pi\omega_2)\sin(\pi\omega_3)}\right]$ & $\tau_1\tau_4$\\
\vspace{0.3cm}
&$e^{i\pi\left(\omega_a-\omega_b+\omega_c\right)}e^{-i\pi\left(\omega_1+\omega_4\right)}$ & $\tau_2\tau_3$\\
\vspace{0.3cm}
&$e^{-i\pi\left(\omega_1+\omega_3\right)}e^{i\pi\omega_a}\left[e^{i\pi\left(\omega_b+\omega_c\right)} - 2ie^{i\pi\omega_3}\frac{\sin(\pi\omega_b)\sin(\pi\omega_c)}{\sin(\pi\omega_3)} \right]$ & $\tau_2\tau_4$\\
\vspace{0.3cm}
&$e^{i\pi\left(\omega_a+\omega_b-\omega_c\right)}e^{-i\pi\left(\omega_1+\omega_2\right)}$ & $\tau_3\tau_4$\\
\vspace{0.3cm}
&$-e^{-i\pi\omega_4}e^{i\pi\omega_c}\left[e^{-i\pi\left(\omega_a+\omega_b\right)} + 2ie^{-i\pi\omega_2}\frac{\sin(\pi\omega_a)\sin(\pi\omega_b)}{\sin(\pi\omega_2)} \right]$ & $\tau_1\tau_2\tau_3$\\
\vspace{0.3cm}
&$-\left[e^{i\pi\left(-\omega_a+\omega_b+\omega_c\right)}e^{-i\pi\omega_3} - 2i\frac{\sin(\pi\omega_{2a})\sin(\pi\omega_b)\sin(\pi\omega_c)}{\sin(\pi\omega_2)\sin(\pi\omega_3)} \right]$ & $\tau_1\tau_2\tau_4$\\
\vspace{0.3cm}
&$-\left[e^{i\pi\left(\omega_a+\omega_b-\omega_c\right)}e^{-i\pi\omega_2} - 2i\frac{\sin(\pi\omega_a)\sin(\pi\omega_b)\sin(\pi\omega_{3c})}{\sin(\pi\omega_2)\sin(\pi\omega_3)} \right]$ & $\tau_1\tau_3\tau_4$\\
\vspace{0.3cm}
&$-e^{-i\pi\omega_1}e^{i\pi\omega_a}\left[e^{-i\pi\left(\omega_b+\omega_c\right)} + 2ie^{-i\pi\omega_3}\frac{\sin(\pi\omega_b)\sin(\pi\omega_c)}{\sin(\pi\omega_3)} \right]$ & $\tau_2\tau_3\tau_4$\\
\vspace{0.3cm}
&$\left[e^{i\pi\left(-\omega_a+\omega_b-\omega_c\right)} - 2i\frac{\sin(\pi\omega_{2a})\sin(\pi\omega_b)\sin(\pi\omega_{3c})}{\sin(\pi\omega_2)\sin(\pi\omega_3)} \right]$ & $\tau_1\tau_2\tau_3\tau_4$.\\
\vspace{0.3cm}
\end{tabular} 
\end{center}
\caption{All terms of the production amplitude $P_{2\rightarrow5}$.}\label{table3}
\end{table}

%%%%%%%%%%%%%%%%%%%%%%%%%%%%%%%%%%%%%%%%%%%%%%%%%%%%%%%%%%%%%%%%%%%%%%%%%%%%%%%%%%%%%%%%%%%%%%%%%%%%
%%%%%%%%%%%%%%%%%%%%%%%%%%%%%%%%%%%%%%%%%%%%%%%%%%%%%%%%%%%%%%%%%%%%%%%%%%%%%%%%%%%%%%%%%%%%%%%%%%%%
%\newpage
\section{Recurrence relations for the coefficients of the expansion in the Regge framework}

Consider a configuration of $k$ crosses ("twists") on the left side and one cross on the right side $n$ (Fig.\ref{reccurences})

\begin{eqnarray}
(-1)^k\tau_{i_1}\tau_{i_2}...\tau_{i_k}\left\{A^n_{i_{1}i_{2}...i_{k}}\right\}\;\;\;\;\;\text{and}\;\;\;(-1)^{k+1}\tau_{i_1}\tau_{i_2}...\tau_{i_k}\tau_n\left\{B^n_{i_{1}i_{2}...i_{k}}\right\}\;\;\;;\;\;(n>i_k+1)\nonumber\\
\end{eqnarray}

\begin{figure}[H]
\centering
\epsfig{file=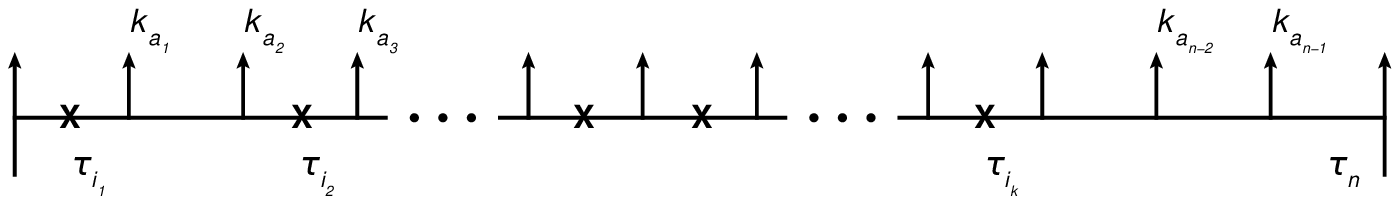,scale=0.9}
%\includegraphics[scale=0.9]{reccurence}
%\caption{Diagram which corresponds to $A^n_{i_{1}i_{2}...i_{k}$ configuration}
%\label{config_1n}
\end{figure}

\begin{figure}[H]
\centering
\epsfig{file=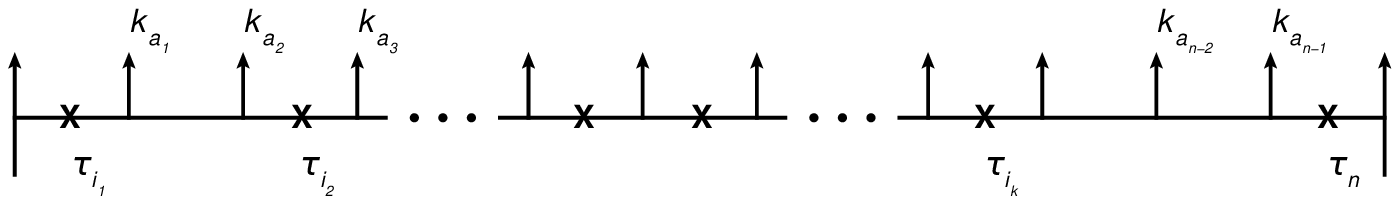,scale=0.9}
\caption{Diagrams which correspond to $A^n_{i_{1}i_{2}...i_{k}}$ (up) and $B^n_{i_{1}i_{2}...i_{k}}$ (bottom) configuration.}
\label{reccurences}
\end{figure}
\noindent
where the recurrence relation reads as
\begin{eqnarray}\label{recrelSimple}
B^{n+1}_{i_{1}...i_{k}}\;=\;B^{n}_{i_{1}...i_{k}}\frac{sin(\pi\omega_{a_n})}{\sin(\pi\omega_n)}+A^{n}_{i_{1}...i_{k}}\frac{sin(\pi\omega_n-\pi\omega_{a_n})}{\sin(\pi\omega_n)}
\end{eqnarray}
This [Eq.\eqref{recrelSimple}] can be rewritten, using Eq.\eqref{identity},
\begin{eqnarray}
B^{n}_{i_{1}...i_{k}}\;=\;b^n_{i_{1}...i_{k}}+a^n_{i_{1}...i_{k}}
\end{eqnarray}
with
\begin{eqnarray}
A^{n}_{i_{1}...i_{k}}\;=\;e^{-i\pi\omega_n}a^n_{i_{1}...i_{k}}
\end{eqnarray}
as
\begin{eqnarray}
b^{n+1}_{i_{1}...i_{k}}\;=\;b^n_{i_{1}...i_{k}}\frac{sin(\pi\omega_{a_n})}{\sin(\pi\omega_n)}\;\;\;\text{and}\;\;\;a^{n+1}_{i_{1}...i_{k}}\;=\;a^n_{i_{1}...i_{k}}e^{-i\pi\omega_n}e^{i\pi\omega_{a_n}}
\end{eqnarray}
with initial conditions
\begin{eqnarray}
A^{i_k+1}_{i_{1}...i_{k}}\;=\;e^{-i\pi\omega_{i_k+1}}e^{i\pi\omega_{a_{i_k}}}A^{i_k}_{i_{1}...i_{k}}\;\;\;\text{and}\;\;\;B^{i_k+1}_{i_{1}...i_{k}}\;=\;b^{i_k+1}_{i_{1}...i_{k}}+e^{i\pi\omega_{i_k+1}}A^{i_k+1}_{i_{1}...i_{k}}.
\end{eqnarray}
Let us generalize for the case $i_k<n-2$, $n+1$ produced particles. $$(-1)^k\tau_{i_1}\tau_{i_2}...\tau_{i_k}\tau_{n-1}\tau_{n}\tilde{B}^n_{i_1 i_2...i_k}$$ From the recurrence relation we have

\begin{equation}
\tilde{B}^{n+1}_{i_1 i_2...i_k}\;=\;\frac{sin(\pi\omega_{a_n})}{\sin(\pi\omega_n)}A^{n}_{i_{1}...i_{k}}+\frac{sin(\pi\omega_n-\pi\omega_{a_n})}{\sin(\pi\omega_n)}B^{n}_{i_{1}...i_{k}}
\end{equation}
with 
\begin{eqnarray}\label{b8}
B^n_{i_1 i_2...i_k}\;&=&\;b^n_{i_1 i_2...i_k}+a^n_{i_1 i_2...i_k}\nonumber\\
A^n_{i_1 i_2...i_k}\;&=&\;e^{-i\pi\omega_n}a^n_{i_1 i_2...i_k}\nonumber\\
a^n_{i_1 i_2...i_k}\;&=&\;e^{-i\pi\omega_n}e^{i\pi\omega_{a_n}}a^n_{i_1 i_2...i_k}.
\end{eqnarray}

\begin{figure}[H]
\centering
\epsfig{file=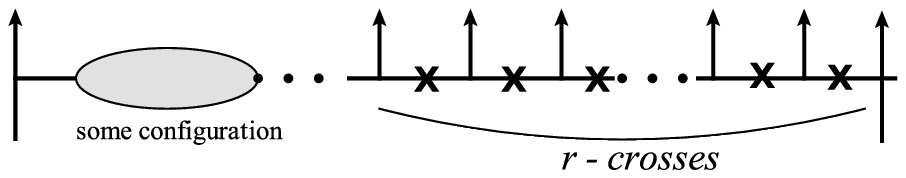,scale=0.9}
\caption{Configuration with $r$ crosses on the most right-hand side and some arbitrary configuration on the left (grey blob),}
\label{gen_config}
\end{figure}

We obtain using the ansatz,
\begin{eqnarray}\label{b9}
\tilde{B}^n_{i_1 i_2...i_k}\;=\;\tilde{b}^{n+1}_{i_1 i_2...i_k}+\tilde{a}^n_{i_1 i_2...i_k},
\end{eqnarray} 
with 
\begin{eqnarray}
\tilde{a}^n_{i_1 i_2...i_k}\;=\;e^{i\pi\omega_n}a^n_{i_1 i_2...i_k}.
\end{eqnarray}
The result is
\begin{eqnarray}
\tilde{b}^n_{i_1 i_2...i_k}\;&=&\;\frac{\sin(\pi\omega_n-\pi\omega_{a_n})}{\sin(\pi\omega_n}b^n_{i_1 i_2...i_k}\nonumber\\
\tilde{a}^{n+1}_{i_1 i_2...i_k}\;&=&\;e^{-i\pi\omega_{a_n}}\;a^n_{i_1 i_2...i_k}\;=\;e^{-i\pi\omega_n}e^{-i\pi\omega_{a_n}}\tilde{a}^n_{i_1 i_2...i_k}.
\end{eqnarray}
Now we consider the most general case, Fig.\ref{gen_config}.
\noindent
\begin{eqnarray}
(-1)^{k+r}\tau_{i_1}\tau_{i_2}...\tau_{i_k}\tau_{n-r}\tau_{n-r+1}...\tau_{n}B^n_r\;\;\;\text{for}\;\;i_k<n-r-1.
\end{eqnarray}
Then the recurrence relation becomes the three-term relation for $B^n_r$:
\begin{eqnarray}
B^{n+1}_{r+1}\;=\;e^{-i\pi(\omega_n-\omega_{a_{n-1}})}\frac{\sin(\pi\omega_{a_n})}{\sin(\pi\omega_n)}B^{n-1}_{r-1}+\frac{\sin(\pi\omega_n-\pi\omega_{a_n})}{\sin(\pi\omega_n)}B^n_r,
\end{eqnarray}
with the initial conditions,
\begin{eqnarray}
B^n_1\;=\;B^n_{i_1 i_2...i_k}\;\;\;\text{and}\;\;\;B^n_2\;=\;\tilde{B}^n_{i_1 i_2...i_k}.
\end{eqnarray}
The recurrence relation for $B^n_{i_1 i_2...i_k}$ Eq.\eqref{b8} and $\tilde{B}^n_{i_1 i_2...i_k}$ Eq.\eqref{b9} are two-term recurrence relations and therefore, we can construct everything in terms of very simple relations. Consider a case $(-1)^{k+r}\tau_{i_1}\tau_{i_2}...\tau_{i_k}..\tau_{n-r}\tau_{n-r+1}...\tau_{n}B^n_{r}$, where we obtain the recurrence relation
\begin{eqnarray}
B^{n+1}_{r+1}\;=\;\frac{\sin(\pi\omega_{a_n})}{\sin(\pi\omega_n)}e^{-i\pi\omega_n}e^{-i\pi\omega_{a_{n-1}}}B^{n-1}_{r-1}+\frac{\sin(\pi\omega_{n}-\pi\omega_{a_n})}{\sin(\pi\omega_n)}B^n_r.
\end{eqnarray}
In particular, 
\begin{eqnarray}
B^{n+2-r}_2\;=\;\frac{\sin(\pi\omega_{a_{n+1-r}})}{\sin(\pi\omega_{n+1-r})}e^{-i\pi\omega_{n+1-r}}e^{-i\pi\omega_{a_{n-r}}}B^{n-r}_0 + \frac{\sin(\pi\omega_{n+1-r}-\pi\omega_{a_{n+1-r}})}{\sin(\pi\omega_{n+1-r})}B^{n+1-r}_1\nonumber\\,
\end{eqnarray}
where we can write
\begin{eqnarray}
B_2\;=\;b_2+a_2;\;\;\;B^n_1\;=\;b^n_1+a^n_1;\;\;\;a_1\;=\;e^{i\pi\omega_{a_{n-1}}}B^{n-1}_0;\;\;\;b^{n+1}_2\;=\;\frac{\sin(\pi\omega_n-\pi\omega_{a_n})}{\sin(\pi\omega_n)}b^n_1\nonumber\\
\end{eqnarray}
and 
\begin{eqnarray}
b^{r+1}_1\;=\;b^2_1\frac{\sin(\pi\omega_{a_{r}})}{\sin(\pi\omega_r)};\;\;\;a^{r+1}_1\;=\;a^r_1e^{-i\pi\omega_r}e^{i\pi\omega_{a_r}}.
\end{eqnarray}
We have
\begin{eqnarray}
B^{n+1}_1\;=\;\frac{\sin(\pi\omega_{a_n})}{\sin(\pi\omega_n)}B^n_1 + \frac{\sin(\pi\omega_n  - \pi\omega_{a_n})}{\sin(\pi\omega_n)}A^n_1;\;\;\;\;\text{with}\;\;\;A^n_1\;=\;e^{-i\pi\omega_n}a^n_1.
\end{eqnarray}
Indeed, 
\begin{eqnarray}
a^{n+1}_1\;=\;\left[\frac{\sin(\pi\omega_{a_n})+e^{-i\pi\omega_n}\sin(\pi\omega_n-\pi\omega_{a_n})}{\sin(\pi\omega_n)}\right]a^n_1\;=\;e^{-i\pi\omega_n}e^{i\pi\omega_{a_n}}a^n_1.\nonumber\\
\end{eqnarray}
Concluding this part, one can see a clear recurrence relation for an arbitrary number of crosses and "holes" (untwisted propagators). Thus, more complicated configurations might be reduced to the more simple ones using the recurrence relations formulated in the above.

\bibliography{biblioFull}{}

\begin{thebibliography}{10}

\bibitem{Bern:2005iz}
Z.~Bern, L.~J. Dixon, and V.~A. Smirnov, ``{Iteration of planar amplitudes in
  maximally supersymmetric Yang-Mills theory at three loops and beyond},'' {\em
  Phys.Rev.}, vol.~D72, p.~085001, 2005.

\bibitem{Alday:2007he}
L.~F. Alday and J.~Maldacena, ``{Comments on gluon scattering amplitudes via
  AdS/CFT},'' {\em JHEP}, vol.~0711, p.~068, 2007.

\bibitem{Bartels:2008ce}
J.~Bartels, L.~Lipatov, and A.~Sabio~Vera, ``{BFKL Pomeron, Reggeized gluons
  and Bern-Dixon-Smirnov amplitudes},'' {\em Phys.Rev.}, vol.~D80, p.~045002,
  2009.

\bibitem{Bartels:2008sc}
J.~Bartels, L.~Lipatov, and A.~Sabio~Vera, ``{N=4 supersymmetric Yang Mills
  scattering amplitudes at high energies: The Regge cut contribution},'' {\em
  Eur.Phys.J.}, vol.~C65, pp.~587--605, 2010.

\bibitem{Lipatov:2009nt}
L.~Lipatov, ``{Integrability of scattering amplitudes in N=4 SUSY},'' {\em
  J.Phys.}, vol.~A42, p.~304020, 2009.

\bibitem{Lipatov:2010qg}
L.~Lipatov and A.~Prygarin, ``{Mandelstam cuts and light-like Wilson loops in
  N=4 SUSY},'' {\em Phys.Rev.}, vol.~D83, p.~045020, 2011.

\bibitem{Lipatov:2010ad}
L.~Lipatov and A.~Prygarin, ``{BFKL approach and six-particle MHV amplitude in
  N=4 super Yang-Mills},'' {\em Phys.Rev.}, vol.~D83, p.~125001, 2011.

\bibitem{Bartels:2010tx}
J.~Bartels, L.~Lipatov, and A.~Prygarin, ``{MHV Amplitude for $3 \to 3$ Gluon
  Scattering in Regge Limit},'' {\em Phys.Lett.}, vol.~B705, pp.~507--512,
  2011.

\bibitem{Bartels:2011nz}
J.~Bartels, L.~Lipatov, and A.~Prygarin, ``{Integrable spin chains and
  scattering amplitudes},'' {\em J.Phys.}, vol.~A44, p.~454013, 2011.

\bibitem{Bartels:2011xy}
J.~Bartels, L.~Lipatov, and A.~Prygarin, ``{Collinear and Regge behavior of $2
  \to 4$ MHV amplitude in N = 4 super Yang-Mills theory},'' 2011.

\bibitem{Prygarin:2011gd}
A.~Prygarin, M.~Spradlin, C.~Vergu, and A.~Volovich, ``{All Two-Loop MHV
  Amplitudes in Multi-Regge Kinematics From Applied Symbology},'' {\em
  Phys.Rev.}, vol.~D85, p.~085019, 2012.

\bibitem{Bartels:2011ge}
J.~Bartels, A.~Kormilitzin, L.~Lipatov, and A.~Prygarin, ``{BFKL approach and
  $2 \to 5$ maximally helicity violating amplitude in ${\cal N}=4$
  super-Yang-Mills theory},'' {\em Phys.Rev.}, vol.~D86, p.~065026, 2012.

\bibitem{Bartels:2014ppa}
J.~Bartels, V.~Schomerus, and M.~Sprenger, ``{Heptagon Amplitude in the
  Multi-Regge Regime},'' 2014.

\bibitem{Hatsuda:2014oza}
Y.~Hatsuda, ``{Wilson loop OPE, analytic continuation and multi-Regge limit},''
  2014.

\bibitem{Basso:2014koa}
B.~Basso, A.~Sever, and P.~Vieira, ``{Space-time S-matrix and Flux-tube
  S-matrix III. The two-particle contributions},'' 2014.

\bibitem{Golden:2014xqa}
J.~Golden, M.~F. Paulos, M.~Spradlin, and A.~Volovich, ``{Cluster
  Polylogarithms for Scattering Amplitudes},'' 2014.

\bibitem{Goncharov:2010jf}
A.~B. Goncharov, M.~Spradlin, C.~Vergu, and A.~Volovich, ``{Classical
  Polylogarithms for Amplitudes and Wilson Loops},'' {\em Phys.Rev.Lett.},
  vol.~105, p.~151605, 2010.

\bibitem{DelDuca:2009au}
V.~Del~Duca, C.~Duhr, and V.~A. Smirnov, ``{An Analytic Result for the Two-Loop
  Hexagon Wilson Loop in N = 4 SYM},'' {\em JHEP}, vol.~1003, p.~099, 2010.

\bibitem{DelDuca:2010zg}
V.~Del~Duca, C.~Duhr, and V.~A. Smirnov, ``{The Two-Loop Hexagon Wilson Loop in
  N = 4 SYM},'' {\em JHEP}, vol.~1005, p.~084, 2010.

\bibitem{Dixon:2011pw}
L.~J. Dixon, J.~M. Drummond, and J.~M. Henn, ``{Bootstrapping the three-loop
  hexagon},'' {\em JHEP}, vol.~1111, p.~023, 2011.

\bibitem{Dixon:2011nj}
L.~J. Dixon, J.~M. Drummond, and J.~M. Henn, ``{Analytic result for the
  two-loop six-point NMHV amplitude in N=4 super Yang-Mills theory},'' {\em
  JHEP}, vol.~1201, p.~024, 2012.

\bibitem{Dixon:2012yy}
L.~J. Dixon, C.~Duhr, and J.~Pennington, ``{Single-valued harmonic
  polylogarithms and the multi-Regge limit},'' {\em JHEP}, vol.~1210, p.~074,
  2012.

\bibitem{Pennington:2012zj}
J.~Pennington, ``{The six-point remainder function to all loop orders in the
  multi-Regge limit},'' {\em JHEP}, vol.~1301, p.~059, 2013.

\bibitem{Dixon:2013eka}
L.~J. Dixon, J.~M. Drummond, M.~von Hippel, and J.~Pennington, ``{Hexagon
  functions and the three-loop remainder function},'' {\em JHEP}, vol.~1312,
  p.~049, 2013.

\bibitem{Lipstein:2013xra}
A.~E. Lipstein and L.~Mason, ``{From dlogs to dilogs; the super Yang-Mills MHV
  amplitude revisited},'' 2013.

\bibitem{Golden:2013xva}
J.~Golden, A.~B. Goncharov, M.~Spradlin, C.~Vergu, and A.~Volovich, ``{Motivic
  Amplitudes and Cluster Coordinates},'' {\em JHEP}, vol.~1401, p.~091, 2014.

\bibitem{DelDuca:2013lma}
V.~Del~Duca, L.~J. Dixon, C.~Duhr, and J.~Pennington, ``{The BFKL equation,
  Mueller-Navelet jets and single-valued harmonic polylogarithms},'' 2013.

\bibitem{Caron-Huot:2013fea}
S.~Caron-Huot, ``{When does the gluon reggeize?},'' 2013.

\bibitem{Lipatov:2010qf}
L.~Lipatov, ``{Analytic properties of high energy production amplitudes in N=4
  SUSY},'' {\em Theor.Math.Phys.}, vol.~170, pp.~166--180, 2012.

\bibitem{Weis:1972ir}
J.~Weis, ``{Factorization of multi-regge amplitudes},'' {\em Phys.Rev.},
  vol.~D4, pp.~1777--1787, 1971.

\bibitem{Weis:1972tn}
J.~Weis, ``{Factorization of multi-regge amplitudes. ii},'' {\em Phys.Rev.},
  vol.~D5, pp.~1043--1047, 1972.

\bibitem{Brower:1974yv}
R.~Brower, C.~E. DeTar, and J.~Weis, ``{Regge Theory for Multiparticle
  Amplitudes},'' {\em Phys.Rept.}, vol.~14, p.~257, 1974.

\bibitem{Drummond:1969ft}
I.~Drummond, P.~Landshoff, and W.~Zakrzewski, ``{The two-reggeon/particle
  coupling},'' {\em Nucl.Phys.}, vol.~B11, pp.~383--405, 1969.

\end{thebibliography}
\bibliographystyle{ieeetr}

\end{document}